\documentclass[epjST]{svjour}
\usepackage{graphics}
\usepackage{lscape}
\usepackage{graphicx}
\graphicspath{ {./images/} }
\usepackage{ulem}
\begin{document}
\title{Ground Based Gamma-ray Astronomy: History and Development of Techniques}
\author{D. Bose\inst{1}\fnmsep\thanks{\email{debanjan.bose@bose.res.in}} \and V. R. Chitnis\inst{2} \and P. Majumdar\inst{3,4} \and B. S. Acharya\inst{2} }
\institute{Department of Astrophysics and Cosmology, S N Bose National Centre for Basic Sciences, Kolkata, India \and Department of High Energy Physics, Tata Institute of Fundamental Research, Mumbai, India \and High Energy Nuclear and Particle Physics Division, Saha Institute of Nuclear Physics, Kolkata, India \and Faculty of Physics and Applied Informatics, Department of Astrophysics, University of Lodz, Poland}
\abstract{
 Very High Energy (VHE) $\gamma$-rays constitute one of the main pillars of high energy astrophysics. Gamma-rays are produced under extreme relativistic conditions in the Universe. VHE $\gamma$-rays can be detected indirectly on the ground. Detection of these energetic photons poses several technological challenges. Firstly, even though $\gamma$-rays are highly penetrative, the Earth's atmosphere is opaque to  them. Secondly, these $\gamma$-rays are to be detected against the overwhelming background of cosmic rays. When a VHE $\gamma$-ray arrives at the top of the atmosphere it produces charged secondaries. These charged particles produce Cherenkov flashes in optical band. Even though first attempts  to detect these Cherenkov flashes were made almost 70 years ago, it took several decades of relentless efforts to streamline the technique. Ground-based VHE $\gamma-$ray astronomy has now established itself as one of the crucial branches of conventional high energy astronomy to study the relativistic Universe. In this article we look back and present a historical perspective followed by a discussion on the current status and finally what lays ahead. 
} 
\maketitle
\label{intro}
\section{Introduction}
Direct detection of $\gamma$-rays originating from celestial sources is only possible 
from space outside the Earth's atmosphere - using either satellite or balloon-borne
detectors. 
The Large Area Telescope (LAT) onboard Fermi $\gamma$-ray Space Telescope \cite{fermi} detects $\gamma$-rays upto energies of few hundreds of GeV following pair-production mechanism. At higher energies the flux of $\gamma$-rays arriving at Earth reduces drastically following a power-law spectrum : $\frac{dN}{dE} \propto E^{-\alpha}$ (where $\alpha$ $\sim$ 2-3 is the spectral index), which means the number of photons detected per unit area per unit time falls rapidly with increasing energy. As a result, beyond a few tens of GeV, statistics is not good enough to do $\gamma$-ray astronomy with space-borne detectors. 
At higher energies, one needs large area detectors and longer exposure times.
Earth's atmosphere absorbs $\gamma$-rays. However, if the $\gamma$-rays are 
energetic enough, with energies above 100 GeV, they interact with the nuclei in the atmosphere 
and  generate a cascade of secondary particles, called Extensive Air Shower (EAS).

Like $\gamma$-rays, cosmic rays also produce EAS when they arrive at the top of the atmosphere. Cosmic rays or Astro-particles were discovered in 1912. By the 1950s much of the properties
of the EAS were understood. But even after more than 100 years of their discovery, the origin
or the sources of cosmic rays are not firmly identified. This is because of the fact 
that cosmic rays being charged particles, their trajectories bend in the magnetic field 
of the source environs and interstellar magnetic field en-route including the magnetic 
field of the Earth. Thus pointing back the direction of cosmic rays to know their source 
is not possible unless they are of extremely high energy (EeV i.e. 10$^{18}$ eV) and of galactic origin. 
Cosmic rays would interact with matter in the source environs and create neutral 
species like $\gamma$-rays and neutrinos, which could come to Earth directly from 
the source location without bending their trajectories or losing their direction. Thus one could 
use these $\gamma$-rays or neutrinos to locate the sources of cosmic rays. In fact, with a desire
to search for sources of cosmic rays, the cosmic ray physicists started the work on ground-based 
$\gamma$-ray astronomy, though detecting the $\gamma$-rays indirectly, well before the 
advent of space-based detectors used for direct detection of $\gamma$-rays. Alternatively,
neutrinos could also point to the source direction. Being weakly interacting particles,
in fact, they could come from farther distances compared to $\gamma$-rays. But,  unlike $\gamma$-rays, they are 
difficult to detect  and need massive detectors like IceCube \cite{icecube}, KM3NeT \cite{km3net}, Hyper-Kamiokande \cite{hyperk} etc.

The secondary particles in the 
EAS, produced by $\gamma$-rays, which are mostly electrons and positrons, also get absorbed in the atmosphere, unless the 
energy of the $\gamma$-ray is high enough ($\sim$ 100 TeV or so) for a sizeable number of charged 
particles to reach the ground level.  In such cases, where the electrons and positrons 
of the EAS reach the ground level, the shower could be detected by an array of particle 
detectors. From the relative arrival times of the shower front at different detectors, one can 
determine the  direction of the shower axis in space by the method of triangulation. 
The shower axis direction is also the arrival direction of the $\gamma$-ray.

The secondary charged particles in the EAS also cause the emission of Cherenkov radiation,
which is beamed in the forward direction and can reach ground level. So, at energies in the
range of a few 10's of GeV to a few 10's of TeV, even though charged particles do not reach
the ground, it is possible to detect Cherenkov light and indirectly detect primary $\gamma$-ray.
This technique, called the atmospheric Cherenkov technique, with the entire atmosphere acting as 
a detection medium is the most efficient way of detecting Very High Energy (VHE) $\gamma$-rays 
spanning an energy range of a few 10's of GeV to a few 10's of TeV. For higher $\gamma$-ray energies, 
above 100 TeV, the sizeable number of charged particles reach the ground and are detected using
charged particle detectors. This is Ultra High Energy (UHE) $\gamma$-ray astronomy.

VHE $\gamma$-ray astronomy activities started in the 1960s and evolved over the
last sixty years. The primary physics objectives of $\gamma$-ray astronomy are
to establish sites and mechanisms for the origin of cosmic rays and their
acceleration to energies way beyond those achieved with man-made accelerators,
thereby solving the 100 years old problem of galactic and extragalactic
cosmic-ray origin. This is to be done by carrying out multiwavelength and multi-messenger observations
of astrophysical sources like supernova remnants, pulsar wind nebulae,
active galactic nuclei etc. Secondly, exploring how transparent is our
Universe i.e. cosmological studies with TeV $\gamma$-rays and understanding
the extragalactic background light leading to an estimation of Hubble
parameter thus aiming to constrain the expansion parameters of the Universe.
Another topic is Indirect Detection of Dark Matter through observations of
dark matter dominated galaxies thus complementing the ongoing efforts through
direct detection techniques and collider searches. Also, there are other
points to investigate like the role of cosmic rays in star forming systems,
understanding the nature of the central engine in very high energy gamma
ray bursts, probing fundamental physics through studies of Lorentz
invariance violation and search for axion-like particles.

In this review, we outline the evolution of the techniques used for the detection of 
VHE $\gamma$-rays, starting from the historical era of the first generation telescopes (1960-1988),  the second generation telescopes 
after a major breakthrough in the field brought by the Whipple telescope, the third generation 
i.e. the present generation telescopes followed by telescopes planned for the future. Even though the main 
emphasis is on the telescopes based on the atmospheric Cherenkov technique, we also cover 
other techniques like air shower arrays which are also
used for detection of VHE $\gamma$-rays. Historical aspects are given in section \ref{hist}. This 
is followed by details of the atmospheric Cherenkov technique in section \ref{act}. The wavefront sampling telescopes and imaging atmospheric Cherenkov telescopes are discussed in sections \ref{wcts} and \ref{iacts}. Simulations and data analysis technique is explained in section \ref{sim}. In section \ref{eas} we have described the air shower array technique and finally conclusions in section \ref{con}.

\section{Historical Background}
\label{hist}

G. Cocconi's proposal in 1960 \cite{Cocconi}, to search for $\gamma$-ray sources by detecting 
charged particles in EAS using an array of scintillators at the mountain 
altitudes could be considered as a starting point for ground-based $\gamma$-ray astronomy. 
Another possibility was to use the particular feature of $\gamma$-ray initiated showers 
like paucity of muons in EAS compared to cosmic ray showers. Later G. T.
Zatsepin \cite{Zatsepin} proposed to use Cherenkov light produced by 
EAS instead of charged particles as proposed by G. Cocconi. The births of VHE and UHE 
branches of $\gamma$-ray astronomy occurred at the same time. 

VHE $\gamma$-ray astronomy based on atmospheric Cherenkov technique owes its existence 
to the pioneering works carried out by W. Galbraith and J. V. Jelley in the early 1950s \cite{Galbraith}. They discovered the presence of Cherenkov light in EAS following a 
suggestion by P. M. S. Blackett to look for Cherenkov light in air \cite{Blackett}. 

A very first large scale experiment with twelve 1.5m diameter parabolic mirrors on the mount 
was carried out by a USSR team led by A. E. Chudakov at Katsiveli, Crimea in 1960-63
\cite{RefCremia} (see Fig. 2 of \cite{Lidvansky} for photograph of the array).  This was followed by a British-Irish experiment at 
Glencullen, Ireland by J. V. Jelley and N. A. Porter, having two 90 cm diameter searchlight mirrors on a Bofors gun mounting  \cite{Jelley-Porter}. The Smithsonian group 
in the USA led by G. Fazio built a 10-m diameter light reflector telescope at Mt. Hopkins at Arizona, 
USA (Fred Lawrence Whipple Observatory) in 1968 \cite{Fazio}.  The early attempts 
were also carried out by various groups in several parts of 
the world viz., USSR, USA, UK, Ireland, Japan, Australia, India, France, Italy and 
South Africa. These first-generation experiments were simple and did not have the 
ability to reject or identify the numerous cosmic ray generated EAS which posed as a 
background in the detection of $\gamma$-rays. Most experiments did not detect any source 
firmly but detected sources occasionally and some detected transient emissions or 
sporadic emissions from sources like Crab, Vela, Geminga, X-ray binaries (Cyg X-3, 
Her X-1, SS-433) and AGNs (Cen-A) etc. None detected consistent steady emission 
from Crab nebula. (For example, see the review article by T. C. Weekes \cite{Weekes_rev_1989} and references therein.)
Thus prospects of ground-based VHE $\gamma$-ray astronomy looked 
bleak due to a very low flux of $\gamma$-rays and high cosmic ray background. Nevertheless, 
the ground-based techniques flourished mainly due to
technical difficulties of detecting higher 
and higher energy photons directly with space-based detectors because of
continuously decreasing $\gamma$-ray 
flux with increasing energy. The effective detection area 
($>10^4~m^2$) offered by the ground-based technique could hardly be realized in detectors 
deployed in space.

There were lots of efforts to reduce the cosmic ray background and thereby improve 
the sensitivity for detecting $\gamma$-rays as well as to reduce the threshold energy
i.e. the lowest $\gamma$-ray energy that can be detected by the experimental system using
larger area light collectors, improved electronics etc ( for a detailed description of sensitivity and energy threshold see section \ref{sim}). The discovery of radio pulsar 
in 1968 \cite{Hewish} offered some method of rejecting a fraction of the cosmic 
ray background from sources like pulsars and X-ray binaries. The arrival times  of 
cosmic rays are random while for $\gamma$-rays from pulsars are modulated 
with the pulsar spin frequency. Similarly, arrival times of $\gamma$-rays from binary 
objects (like Cyg X-3) are modulated with orbital frequency. Efforts were also 
directed toward exploring the subtle differences between cosmic ray initiated EAS 
and $\gamma$-ray initiated ones to reject the vast cosmic ray background. Thus several 
characteristics like distribution of Cherenkov photons in the light pool of the 
shower, the time structure of Cherenkov light, the fraction of UV light content in the 
shower, angular spread of Cherenkov photons within the shower etc. were being 
exploited to reduce the cosmic ray background (see for example, review article
by T. C.  Weekes \cite{Weekes_2005} and references therein). An attempt of imaging the shower 
(albeit a low resolution image) was carried out by D. Hill and N. A. Porter in 1961 using 
image intensifiers following a suggestion from J. V. Jelley in 1958 \cite{Hill}. The 
potential advantage of using shower images for discriminating $\gamma$-ray induced 
showers from those of cosmic rays was listed by J. V. Jelley and  N. A. Porter in a review in 
1963 \cite{Jelley-Porter}. Early efforts in imaging the shower have been described
in an article by N. A. Porter \cite{Porter}.

\section{Atmospheric Cherenkov Technique }
\label{act}

As mentioned earlier, the Earth's atmosphere is opaque to $\gamma$-rays. However, upon arrival at the top of the atmosphere, a VHE $\gamma$-ray  produces an electromagnetic shower. Charged particles present in this shower cause atmosphere to emit Cherenkov radiation. These secondary particles or radiation can be detected at the ground level. Telescopes that detect $\gamma$-rays by detecting Cherenkov radiation produced in the atmosphere are called Atmospheric Cherenkov Telescopes (ACTs).

\begin{figure}%
    \centering
    {{\includegraphics[width=6.0cm]{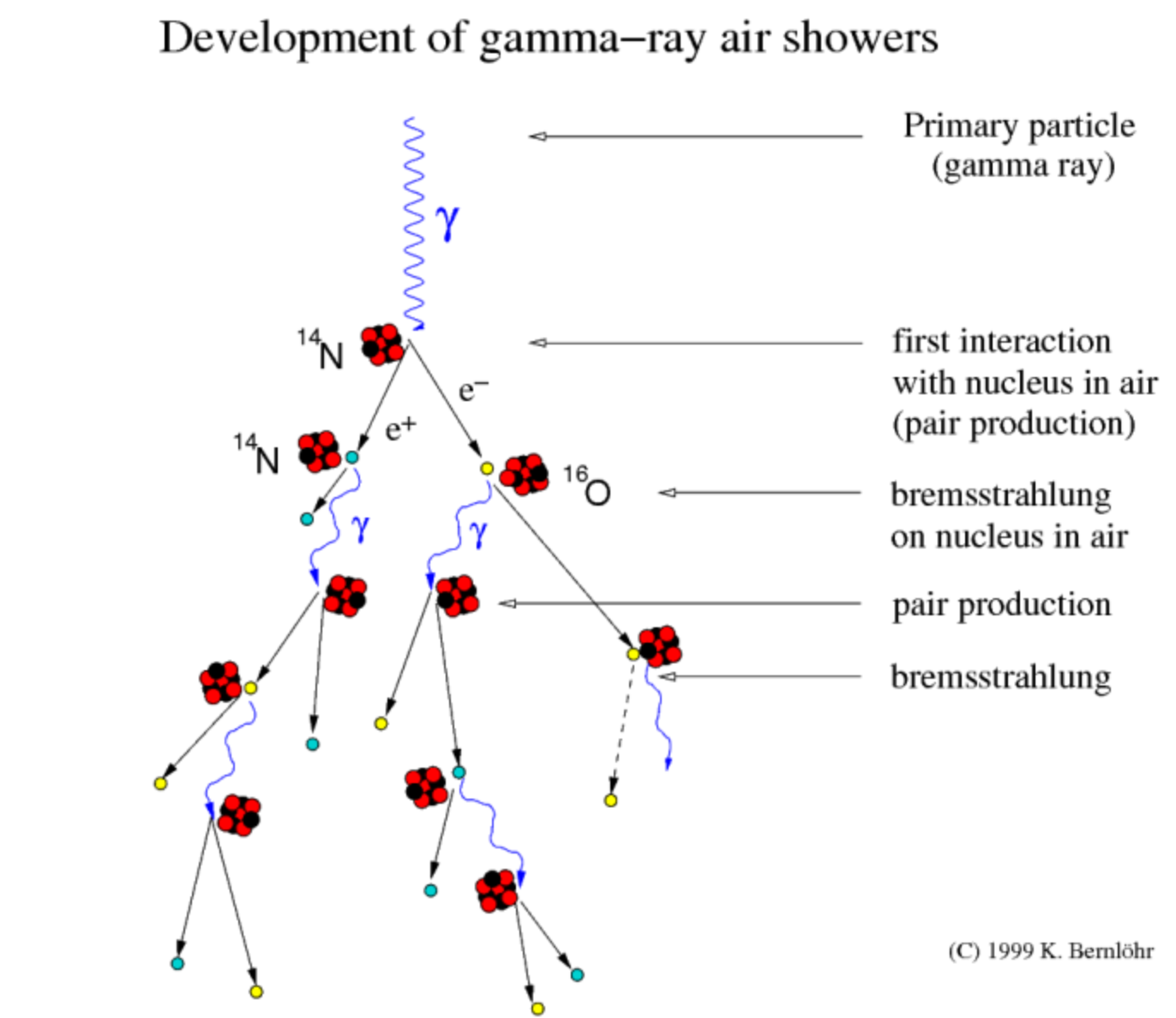} }}%
    \qquad
    {{\includegraphics[width=6.0cm]{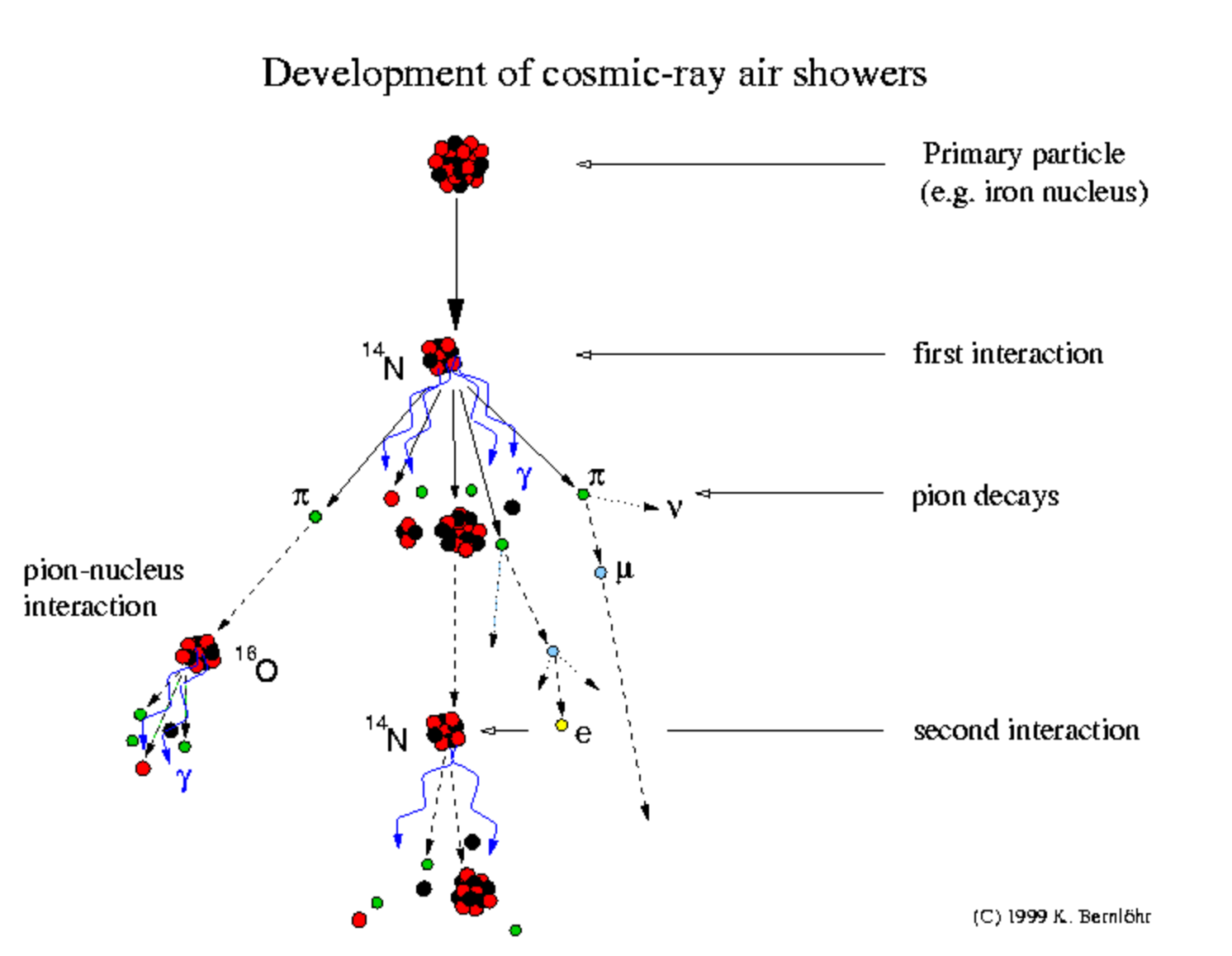} }}%
    \caption{EAS initiated by  a  $\gamma$-ray (left) and  a cosmic ray (right). $\gamma$-ray produces  electromagnetic cascade whereas cosmic ray produces three overlapping cascades: electromagnetic, pionic and nuclear. Reproduced from  https://www.mpi-hd.mpg.de/hfm/CosmicRay/Showers.html with permission. from author Konrad Bernl$\ddot{o}$hr.}%
    \label{fig:em}%
\end{figure}

\begin{figure}[h]
    \centering
    \includegraphics[width=0.65\textwidth]{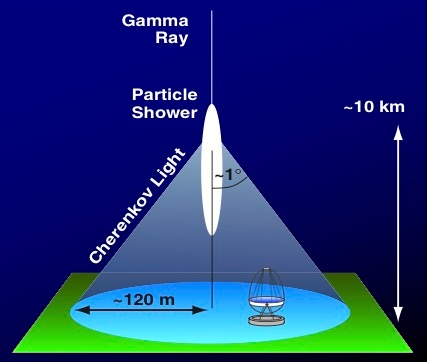}
    \caption{Principle of atmospheric Cherenkov technique. Celestial $\gamma$-ray generates  electromagnetic shower in atmosphere. Secondary charged particles in the shower cause atmosphere to emit Cherenkov radiation in a small cone. Cherenkov photons create a light pool on the ground spread over a circular area of radius $\sim$ 120 m. This light is detected using Atmospheric Cherenkov Telescope (ACT). (Reproduced from \cite{Naurois_Mazin} with permission from authors M. de Naurois and D. Mazin)}
    \label{fig:cherenkov}
\end{figure}

When a $\gamma$-ray enters into the Earth’s atmosphere it produces an $e^{\pm}$ pair. This pair then emits $\gamma$-rays via bremsstrahlung. With successive pair-production and bremss-trahlung, an electromagnetic cascade develops as shown in the figure \ref{fig:em}. Charged particles i.e. $e^{\pm}$ pairs present in this cascade travel with a speed more than the speed of light in the medium and therefore cause emission of  Cherenkov radiation peaking in the UV-blue region of the visible spectrum, lasting for a few ns. This radiation is very faint and emitted in a small angle ($\sim$ 1$^\circ$ in the atmosphere) in the forward direction (see figure \ref{fig:cherenkov}). The forward direction and coherent nature of this radiation enable one to determine the arrival direction of the incident $\gamma$-ray accurately. Cherenkov photons are spread over a large circular area with a diameter in the range of 100  - 120 m on the ground. One or multiple telescopes located anywhere in this light pool can detect the incident $\gamma$-ray indirectly by detecting Cherenkov photons. Hence effective area for these telescopes is of the order of $10^4$ to $10^5$ $m^2$. 
 
These telescopes or ACTs are essentially made up of an optical reflector i.e. a mirror to collect the Cherenkov light and one or more Photo Multiplier Tubes or PMTs at the focus of the reflector.  PMTs are the ideal detectors for this purpose because of their high gain, low noise and ultra-fast (ns) response. Also, their spectral response
matches with the Cherenkov spectrum. These telescopes are equipped with very fast front-end electronics to process electrical signals produced by the PMTs.

There are two types of backgrounds that pose a challenge for ground-based $\gamma$-ray astronomy: EAS generated by cosmic rays (mostly proton primaries) and night sky background (NSB). For every $\gamma$-ray initiated EAS there are about 1000 cosmic ray initiated EAS, which can be detected by ACT. 

Cosmic rays initiate a cascade by collision with atmospheric
nuclei, resulting in the production of short-lived pions which include neutral pions
and charged pions. Neutral pions decay into $\gamma$-rays and charged pions produce muons via decay along with other particles. 
These $\gamma$-rays and decay products of muons constitute an electromagnetic
component of EAS (figure \ref{fig:em}). All these charged particles produce Cherenkov radiation.
So, basically 
mechanisms for emission of Cherenkov light are identical for $\gamma$-ray and cosmic ray initiated EAS. Thus the success of any ground-based telescope hinges upon how well it can distinguish a $\gamma$-ray initiated shower from a cosmic ray or a hadronic shower. Because of the differences in the development of air showers initiated by $\gamma$-ray and cosmic ray,  there are some subtle differences in Cherenkov light distribution produced by them on the ground. $\gamma$-ray showers can be separated from hadronic ones by comparing their Cherenkov light pattern.  


The Cherenkov radiation emitted by secondary charged particles in EAS is very faint compared to NSB. Therefore ACTs can only be operated at a dark site, away from
manmade light pollution, during moonless nights with clear weather. The number of NSB photons is typically of the order of 10$^{12}$ $photons/m^2/s/sr$, several orders of magnitude higher compared to the number of Cherenkov photons. For example, the number of Cherenkov photons produced by a $\gamma$-ray of energy 1 TeV is only of the order of a few hundreds per $m^2$ at 2 km altitude a.s.l. These NSB photons peak at higher wavelengths compared to Cherenkov photons and are incoherent. NSB can be suppressed by several orders by selecting PMTs whose quantum efficiency peaks in the wavelength range 300 nm - 400 nm, selecting a narrow time window ($\sim$ few tens of ns) for charge integration and by allowing coincidences between multiple pixels or telescopes.

There are two types of techniques used to detect a $\gamma$-ray signal from an astrophysical source, namely the wavefront sampling technique and the imaging technique.
The wavefront sampling technique is the older one and has been extensively used in the past. However, in the last 20 years, it has been shown that the 
imaging technique is a more powerful tool to detect VHE $\gamma$-rays on the ground due to superior rejection of cosmic ray produced background. 
In the following sections, both these techniques are discussed in detail.

\subsection{Wavefront Sampling Technique }
\begin{figure}[h]
    \centering
    \includegraphics[width=0.65\textwidth]{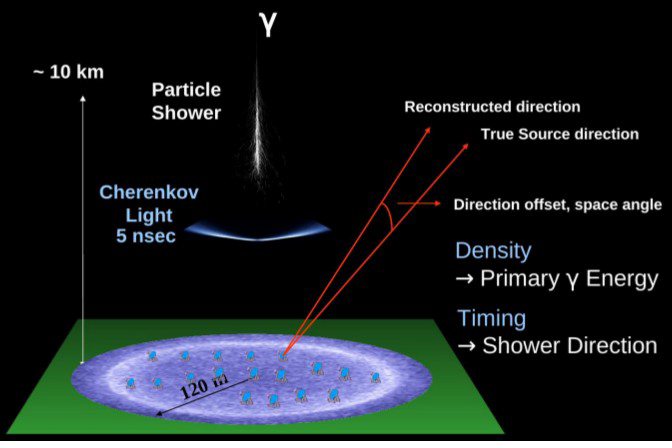}
    \caption{Wavefront sampling technique with multiple collectors sampling light from across Cherenkov pool.}
    \label{fig:ws}
\end{figure}

In the wavefront sampling technique, multiple collectors sample light from across the Cherenkov pool. In this technique, the shower is sampled in the horizontal plane. An array of reflectors is spread across long-baseline with one or more PMTs in the focal plane of each reflector. These reflectors record arrival times 
of Cherenkov shower front and density of photons at several points in the light pool (see figure \ref{fig:ws}). Using arrival time  information, the Cherenkov shower front can be reconstructed. This in turn gives the arrival direction of the incident $\gamma$-ray. The photon density from various PMTs can be used to reconstruct the lateral density profile of Cherenkov light in the shower. This also gives an estimate of the total Cherenkov yield of the shower which is proportional to the energy of the incident $\gamma$-ray primary.

\subsection{Imaging Technique}
The most successful technique to detect Cherenkov light from EAS is to make 
so-called ``pictures" of the showers. The telescopes which employ this technique 
are called Imaging Atmospheric Cherenkov Telescopes (IACTs). These telescopes 
consist of large  reflectors which focus the Cherenkov  light from air 
showers onto a focal plane where a camera is placed. The diameters of the reflectors 
used in various telescopes are in the range of 4m to 28m. The reflector consists
of spherical, hexagonal or square mirror facets made up of glass or aluminium.
The typical size of mirror facets is about 0.5-1 m  and there are about a few hundred mirror 
facets in a medium size (10-12 m diameter) telescope. These mirror facets are
arranged generally in the form of Davies-Cotton or paraboloid design. Davies-Cotton
configuration is simpler in design with identical mirror facets and straightforward
alignment procedure, have inferior on-axis and superior off-axis
focusing. On the other hand, the paraboloid design is slightly difficult to manufacture and
align as all facets have different radii of curvature, has superior on-axis
and inferior off-axis focusing. Davies-Cotton design has another drawback of the
requirement of longer integration times for images as light from various parts
of the reflector reaches the focal plane at different times. Whereas for paraboloid
design, light from different parts of the reflector reaches the focal plane
simultaneously, thereby reducing integration time for images. So typically
Davies-Cotton design is common for small or medium-size telescopes upto about
12m diameter, whereas larger telescopes are mostly based on paraboloid design.

\begin{figure}[h]
    \centering
    \includegraphics[width=0.65\textwidth]{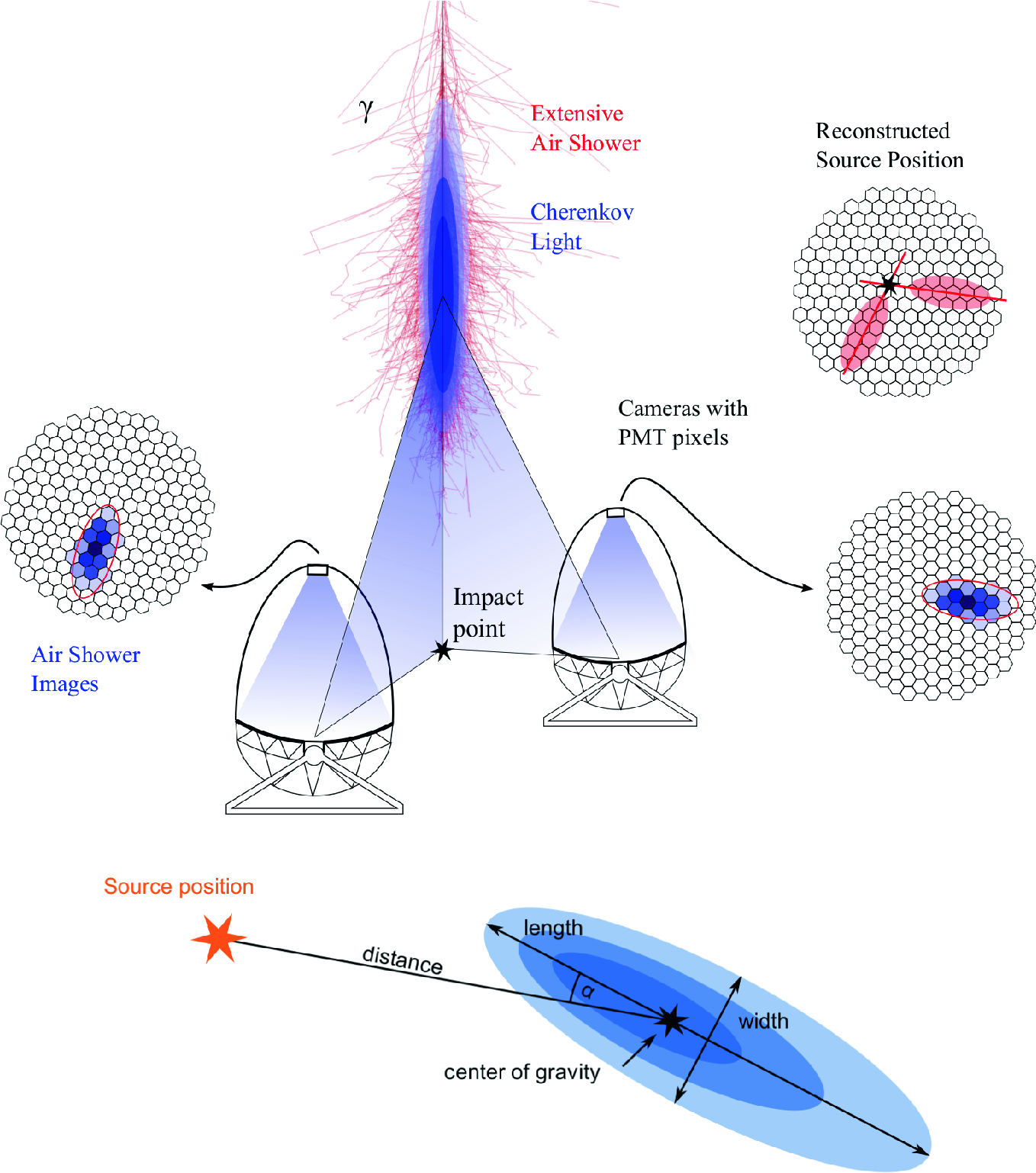}
    \caption{A schematic view of stereoscopic observations of two imaging telescopes. The lower figure
    shows the shower image parameters based on a moment analysis. The image of the shower is fitted
    by an ellipse where the semi-major and semi-minor axes are the length and width parameters
    and represent the ‘shape’ of the image. The alpha parameter is used to estimate the orientation or the pointing 
    of the image. The authors acknowledge the MAGIC collaboration for providing this diagram.}
    \label{fig:imaging}
\end{figure}

The camera in the focal plane of IACT comprises many PMTs, which act 
as pixels. A typical field of view of a camera is 3$^\circ$-5$^\circ$ as the size of
Cherenkov image is typically around a degree. Normally light concentrators
are mounted in front of each PMT pixel to  increase collection area
by taking care of dead space between the PMTs as well as to cut off stray
light falling on the camera. Typical pixel size varies in the range of 
0.1$^\circ$-0.3$^\circ$ and typically there are 500-1000 pixels in the camera of 
medium size telescope with a diameter of $\sim$ 12 m. The camera records an image 
of the shower, and several properties of the image, for example, its shape, 
intensity and orientation, allow one to determine the properties of the 
shower primary (see figure \ref{fig:imaging}). 

\section{Wavefront Sampling Telescopes}
\label{wcts}
As mentioned earlier, the wavefront sampling technique is an older technique used for the detection of Cherenkov photons. Some of the examples from past generation experiments based on this technique are the Pachmarhi Array of Cherenkov Telescopes (PACT) and the  Tracking High Energy Muons In Showers Triggered On Cherenkov Light Emission (THEMISTOCLE) experiment. PACT was a multi-telescope array in central India at an altitude of 1075 m \cite{Bose}. There were 24 telescopes spread over an area of 80 m $\times$ 100 m. Each telescope  consisted of seven 90 cm diameter parabolic mirrors with PMT mounted at the focal point of each mirror. THEMISTOCLE experiment was located at  THEMIS solar power plant site near Font-Romeu in the eastern French Pyrenees \cite{Baillon}. There were 18 telescopes, each having an 80 cm diameter parabolic mirror and a single PMT at the focus. These telescopes were distributed over the roughly elliptical 200 m $\times$ 300 m area. Another array of seven larger, 7 m diameter reflectors, called AStronomie GAmma a Themis (ASGAT), was also operated near THEMISTOCLE \cite{Basiuk}. All these experiments had energy thresholds close to a few hundred GeV or higher, (PACT : $\sim$ 750 GeV, THEMISTOCLE : $\sim$ 3 TeV, ASGAT : $\sim$ 500 GeV). 

\begin{figure}[h]
    \centering
    \includegraphics[width=0.65\textwidth]{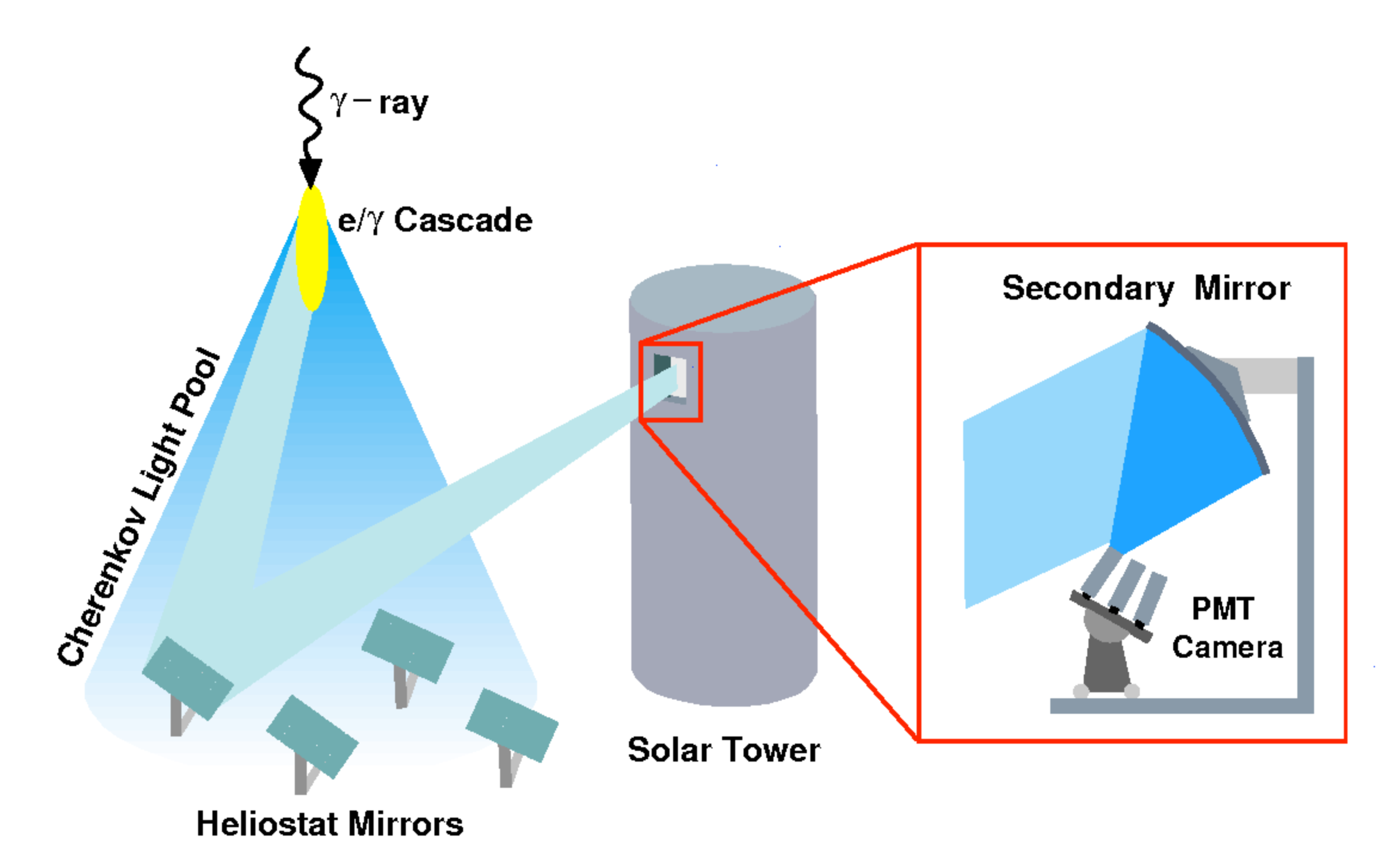}
    \caption{Solar array experiments where Cherenkov photons are reflected by heliostat mirrors towards secondary mirrors located inside the tower. These secondary mirrors focus  Cherenkov light onto an array of PMTs. Image is re-produced with permission from \cite{stacee-img}.}
    \label{fig:solar}
\end{figure}

Around the same time few solar array experiments like Solar Tower Atmospheric Cherenkov Effect Experiment (STACEE) \cite{Gingrich}, Cherenkov Low Energy Sampling and Timing Experiment (CELESTE) \cite{Pare} and Gamma-Ray Astronomy at ALmeria (GRAAL) \cite{Arqueros} were built with the aim of reducing  energy thresholds to bridge the gap between space-based and ground-based detectors. All these experiments used 
large heliostats of solar power facilities to collect more Cherenkov photons to reduce the energy threshold. Light from heliostats was reflected onto secondary mirrors located inside a tower, which then focused light onto a camera consisting of multiple PMTs (see figure \ref{fig:solar}). STACEE was located in 
Albuquerque, New Mexico, USA, CELESTE in the French Pyrenees and GRAAL was located in Almeria, Spain. Due to the larger mirror area, these arrays could achieve energy thresholds in the range of 60-250 GeV.  Details of these 
wavefront sampling arrays are given in Table \ref{tab:ws}.

As mentioned earlier, cosmic rays form a huge background against which $\gamma-$ray signal is to be detected. This is achieved in the wavefront sampling technique
by rejecting off-axis showers which are produced by cosmic rays. Apart from
this, some Gamma-Hadron Segregation (GHS) criteria are also used. For example,
higher fluctuations in cosmic ray showers compared to $\gamma$-rays
arising from shower kinematics were used by CELESTE, to get moderate
rejection \cite{Naurois}. Slightly better rejection was obtained by CELESTE and STACEE
using the shapes of the pulses recorded by flash-ADCs. In this method, pulses from various telescopes were shifted and 
added together after considering arrival time delays. 
A parameter that defines the ratio of height and width of the summed signal was used for
selecting $\gamma$-ray showers (H/W) as distribution of H/W is narrower for $\gamma-$rays compared to cosmic rays \cite{Bruel}. 
Sources like Crab nebula, which is also considered as a standard candle in VHE $\gamma-$ray astronomy, blazar class Active Galactic Nuclei (AGNs) 
like Mrk 421 were successfully detected by these arrays \cite{Naurois,DSmith,Oser,Boone}.
Overall, it turned out that the GHS criteria used in the wavefront sampling technique are less efficient
compared to the imaging technique as we will see later in this review. Hence slowly
imaging technique became more popular and is currently the most advanced and preferred option to detect VHE $\gamma$-rays from celestial sources. 

\begin{figure}[h]
    \centering
    \includegraphics[width=0.65\textwidth]{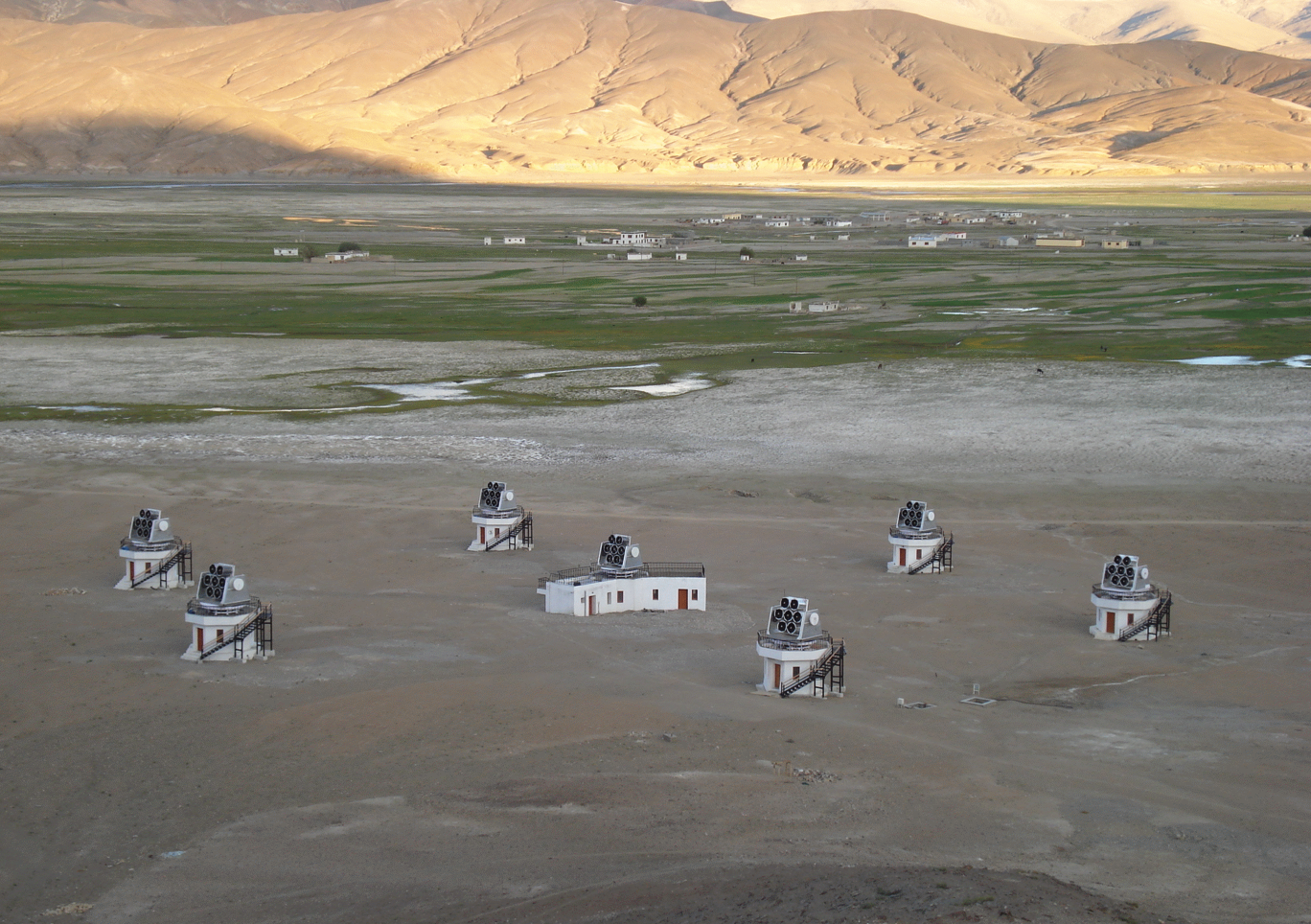}
    \caption{7-element HAGAR telescope array located at Hanle in India at an altitude of $\sim$ 4.3 kms asl. Photo credit : HAGAR group}
    \label{fig:hagar}
\end{figure}

Presently, only one wavefront sampling array, High Altitude GAmma-Ray (HAGAR) telescope system is operational. It is an array of seven small size telescopes, operated at a high altitude location called Hanle in the Ladakh region of the Himalayas in India since 2009 \cite{Saha}. Each of these telescopes has seven para-axially mounted parabolic mirrors of diameter 0.9 m (see figure \ref{fig:hagar}).
 The altitude of Hanle is
4270 m a.s.l. and HAGAR is the first atmospheric Cherenkov telescope to be
operated at such a high altitude location. High altitude location was
chosen as a cost effective way to achieve a lower energy threshold with
a moderate size telescope. The energy threshold depends on the number of
Cherenkov photons collected by the reflector. So one way of reducing the energy threshold is to use a large
size reflector. Whereas alternative way is to install a telescope at high
altitude location. Cherenkov photon density is higher at higher altitudes
and atmospheric absorption of Cherenkov photons is less, which
reduces the energy threshold of the telescope. Taking advantage of high
altitude location, HAGAR could achieve the energy threshold of $\sim$ 210 GeV.
This is about a factor of four reduction compared to similar (in fact,
larger) array PACT which was operated at an altitude of 1 km.

\section{Imaging Atmospheric Cherenkov Telescopes}
\label{iacts}
 The imaging technique has been in use for more than thirty years and has evolved with time. In this section, we summarise the past, present and future generation 
IACTs. 

\subsection{Past generation IACTs}

\begin{figure}[h]
    \centering
    \includegraphics[width=0.65\textwidth]{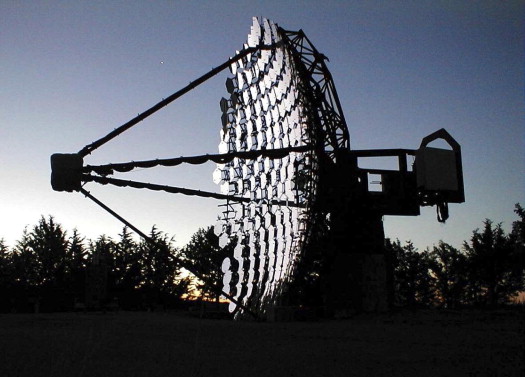}
    \caption{Whipple 10m imaging atmospheric Cherenkov telescope. Reproduced with permission from \cite{whipple-img}. }
    \label{fig:whip}
\end{figure}

The observations of very high energy gamma rays using the imaging technique was pioneered by the 
Whipple telescope collaboration. This is the
same 10 m diameter reflector used by the group led by G. Fazio earlier.
It was located at Mount Hopkins, Arizona at an altitude of 2300m a.s.l.
The reflector was composed of 248 hexagonal aluminized and anodized glass 
mirror facets mounted in Davies-Cotton design (see Fig. ~\ref{fig:whip}). 
Mirror facets had radius 
of curvature of 14.3 m and were mounted on a 7.3 m radius spherical
support structure. Based on the idea of using matrices of PMTs as a
camera to image Cherenkov light from air showers from T. C. Weekes and 
K. E. Turver \cite{Weekes1977}, a 37-pixel camera was installed in the focal 
plane of the reflector. Through Monte Carlo simulations A. M. Hillas 
identified so-called Hillas parameters, which are image parameters,
to discriminate $\gamma$-rays from huge cosmic ray background \cite{Hillas}. 
Using these GHS parameters, 98\% of the 
background was rejected and Crab nebula was detected at a statistical 
significance of 9 $\sigma$ at energies above 700 GeV \cite{Weekes_crab}. 
Whipple continued operation till 2006. Over the years, imaging camera
went through several transformations with a steady increase in the number of 
pixels. In the final configuration, there were 379 pixels. Apart from
Crab nebula, Whipple discovered VHE $\gamma$-ray emission from four blazar
class AGNs : Markarian 421 in the year 1992 \cite{Punch}, 
Markarian 501 in  1996 \cite{Quinn}, 1ES 2344+514 in 1998 \cite{Catanese} and H1426+428 in  2002 \cite{Petry}.

Few years after Whipple discovered emission from Crab nebula, another
IACT called Cherenkov Array at Themis (CAT) was built in France. Though,
using a smaller reflector compared to Whipple, with a diameter of about 5 m,
CAT could achieve performance comparable to that of Whipple because of
finer pixel camera consisting of 546 PMT pixels of 0.12$^\circ$ size
and 54 larger PMTs in two guard rings, covering a total FOV of 4.8$^\circ$
\cite{Barrau}.

The next milestone in the atmospheric Cherenkov technique was achieved by extending 
the technique to an array of telescopes. This is called "stereoscopy". 
This approach was demonstrated by  
High Energy Gamma-Ray Astronomy (HEGRA) experiment located at La
Palma in Canary Islands at an altitude of 2200 m a.s.l. This air shower
array also had IACT system embedded, consisting of an array of five
telescopes. Telescope installation started in 1992 and went through
various stages including refurbishing and by September 1998 entire
system consisting of five telescopes was operational \cite{Puhlhofer}. 
Each of these
telescopes had 8.5 m$^2$ mirror reflector area and camera consisting
of 271 PMT pixels. Telescopes were arranged at the corners of a square with
a side length of 100 m and one telescope at the centre. HEGRA pioneered the
stereoscopic observation technique providing multiple images of a given
air shower from various directions. This provided a better reconstruction
of shower direction and shower core position, thereby improving angular
resolution, GHS and energy resolution. HEGRA
telescope system was operational till September 2002 and apart from
successfully detecting VHE $\gamma$-ray emission from known VHE emitters 
like Crab nebula \cite{Konopelko_crab} and blazars including Mrk 421, Mrk 501, 1ES2344+514,
H1426+428, it discovered sources like Cas A (shell type supernova remnant) \cite{Aharonian_casa}, 
TeV J2032+4130 (a binary system) \cite{Aharonian_2032} and M 87 (radio galaxy) \cite{Aharonian_M87}.


\begin{landscape}
\begin{table}
\begin{tabular}{|c|c|c|c|c|c|c|c|c|c|c|}
\hline
\textbf{Name}& \textbf{Place} & \textbf{Altitude} &  \textbf{Array} &   \textbf{Energy} & \textbf{Duration}  \\

  &  &  & \textbf{Type} &  \textbf{Threshold} &\textbf{Start - End} \\
\hline
ASGAT & French Pyrenees, France & 1650 m & 7 telescopes & 500 GeV & 1989-1995  \\
THEMISTOCLE & French Pyrenees, France & 1650m & 18 Telescopes & 3 TeV & 1990 - 1995 \\
PACT & Pachmarhi, India & 1075m & 24 telescopes & 750 GeV & 2000 - 2012 \\
CELESTE & French Pyrenees, France & 1650m & Heliostat Array 53 telescopes & 60 GeV & 1999 - 2004 \\
STACEE & Albuquerque, New Mexico, USA & 1705m & Heliostat Array 64 telescopes & 150 GeV & 2001 - 2007 \\
GRAAL & Almeria, Spain & 505m & Heliostat Array 63 telescopes & 250 GeV & 1998- 2001 \\
HAGAR & Hanle, India & 4270m & 7 telescopes & 210 GeV & 2008 - ..\\
\hline
\end{tabular}
\caption{{\label{tab:ws}}List of telescopes designed based on Wavefront Sampling}
\end{table}

\begin{table}
\begin{tabular}{|c|c|c|c|c|c|c|c|c|c|c|}
\hline
\textbf{Name}& \textbf{Place} & \textbf{Altitude} &  \textbf{No. of} & \textbf{Mirror}  &  \textbf{No. of}  &  \textbf{F.O.V.} &  \textbf{Energy} & \textbf{Duration}  \\

  &  &  & \textbf{Telescopes} & \textbf{Size} &  \textbf{Pixels} &  &  \textbf{Threshold} &\textbf{Start - End} \\
\hline
Whipple & Arizona, USA & 2300m & 1 & 10m & 379 & 2.8$^\circ$ & 300 GeV & 1985 - 2006 \\ 
GT-48 & Crimea, Ukraine & 600 m & 2 & 4 $\times$ 1.2 m & 150 & 2.6$^\circ$ & 1 TeV & 1989 - 2010 \\
SHALON & Tien Shan, Kazakhstan & 3340 m & 2 & 3.6 m & 144 & 8$^\circ$ & 800 GeV & 1992 - .. \\
CAT & French Pyrenees, France & 1650m & 1 & 4.9m & 546 & 4.8$^\circ$ & 250 GeV & 1996 - 2002 \\
7 Telescope Array & Utah, USA & 1600m & 7 & 3m & 256 & 4$^\circ$ & 600 GeV & 1997 - 1999  \\
HEGRA & Canary Islands, Spain & 2200m & 5 & 3.4m & 271 & 4.3$^\circ$ & 500 GeV & 1992 - 2002  \\
CANGAROO I & Woomera, Australia & 160m & 1 & 3.8m & 256 & 2.9$^\circ$ & 2.7 TeV & 1991 - 1998 \\
CANGAROO III & Woomera, Australia & 160m & 4 & 10m & 427 & 4$^\circ$ & 200 GeV & 2004 - 2011 \\
H.E.S.S. I & Khomas highlands, Namibia & 1800m & 4 & 12m & 960 & 5$^\circ$ & 160 GeV & 2003 - ..   \\
H.E.S.S. II & Khomas highlands, Namibia & 1800m & 1 & 28m & 2048 & 3.2 $^\circ$ & 30 GeV & 2012 - ..   \\
VERITAS & Arizona, USA & 1268m & 4 & 12m & 499 & 3.5$^\circ$ & 85 GeV & 2007 - ..  \\
MAGIC & Canary Islands, Spain & 2200m & 2 & 17m & 1039 & 3.5$^\circ$ & 50 GeV & 2004 - ..  \\
TACTIC & Mt Abu, India & 1300m & 4 & 3.6m & 349 & 6$^\circ$ & 850 GeV & 2001 - ..  \\
FACT & Canary Islands, Spain &  2200m & 1 &  3.4m &  1440 &  4.5$^\circ$ & 750 GeV & 2011 - .. \\
MACE & Hanle, India & 4270m & 1 & 21m & 1088 & 4$^\circ$ & 20 GeV & 2021 - ..\\
\hline
\end{tabular}
\caption{{\label{tab:imaging}}List of telescopes designed based on Imaging Technique}
\end{table}
\end{landscape}

\subsection{Present generation IACTs}

After the successful operation of Whipple demonstrating the power of imaging
technique, CAT showing advantages of fine pixellated camera and HEGRA
successfully displaying significant improvements in sensitivity and
primary energy estimation with stereoscopy, various groups moved to next 
generation of
IACTs. The aim was to improve the sensitivity and to
reduce the energy threshold to 100 GeV or less to have overlap with satellite-based 
detectors like Fermi-LAT. These third-generation telescopes combined
all three aspects like large reflector size, camera with fine pixels and
stereoscopic operations.  Four such stereoscopic arrays came into operation
since 2003. These include 4-telescope arrays H.E.S.S., CANGAROO-III and VERITAS
and 2-telescope array MAGIC.

\subsubsection{H.E.S.S.}

\begin{figure}[h]
    \centering
    \includegraphics[width=10cm,height=6cm]{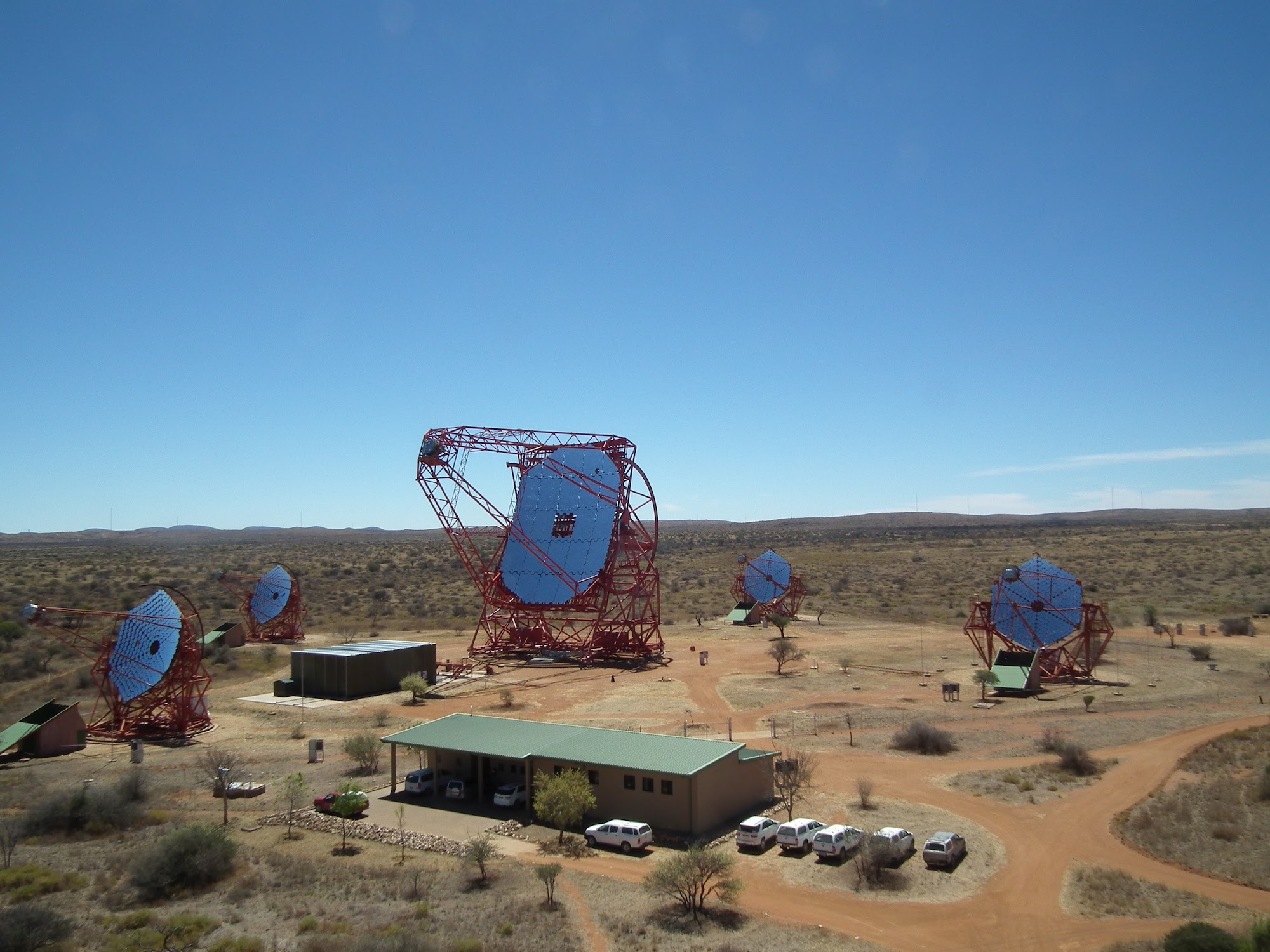}
    \caption{H.E.S.S. IACT array at Gamesberg in Namibia, consisting of four 12 m diameter and one 28 m diameter telescope. Picture courtesy : H.E.S.S. collaboration.}
    \label{fig:hess}
\end{figure}

The High Energy Stereoscopic System\footnote{https://www.mpi-hd.mpg.de/hfm/HESS/} (H.E.S.S.) is an array of four telescopes
located at Gamesberg Mountain in Namibia at an altitude of 1800 m a.s.l. (see fig. \ref{fig:hess}).
Each of these telescopes has a 12m diameter reflector based on
Davies-Cotton design. Each reflector consists of 382 round mirror
facets of 60 cm diameter. The focal length of each telescope is 15 m and
980-pixel camera is mounted in the focal plane. The size of each pixel is
0.16$^\circ$ and the camera covers the field of view of about 5$^\circ$. The angular
resolution of H.E.S.S. is estimated to be 0.1$^\circ$ and the energy threshold
is in the neighbourhood of 100 GeV. H.E.S.S. can detect $\gamma$-ray flux
equivalent to 1\% of Crab in 25 hours of observation \cite{aharonian_2006}. This is almost
an order of magnitude improvement over the previous generation telescopes.

\begin{figure}[h]
    \centering
    \includegraphics[width=0.65\textwidth]{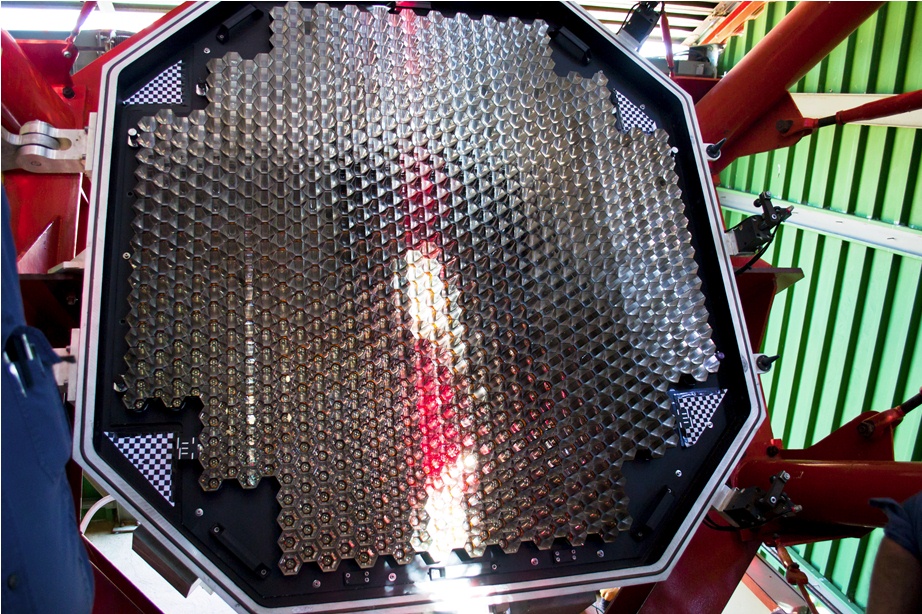}
    \caption{Photograph of the H.E.S.S. II camera, consisting of 2048 pixels. Picture credit : Derek Duckitt, Hermanus Astronomy Centre, South Africa}
    \label{fig:hess_camera}
\end{figure}

H.E.S.S. array is operational since 2003. This was the first telescope
array with good sensitivity operating in the Southern hemisphere. It
carried out the first high resolution and sensitive survey of the
Milky way in TeV $\gamma$-rays \cite{abdalla_2018_milky_way}. It discovered more than 100 objects
of diverse classes. Another interesting measurement from H.E.S.S. is
that of the energy spectrum of cosmic electrons above 600 GeV, taking
advantage of a very large collection area offered by atmospheric
Cherenkov technique compared to direct measurements by satellite
or balloon-borne detectors \cite{aharonian_2008}.

In 2012, the next phase of H.E.S.S., consisting of a single telescope
with a 28m diameter reflector was installed. This is the largest atmospheric
Cherenkov telescope in the world. It has a parabolic reflector and camera
consisting of 2048 pixels (see fig.~\ref{fig:hess_camera} for camera picture). Taking advantage of the lower energy threshold
achieved due to the large reflector size, this telescope could detect
Vela pulsar from sub-20 GeV to 100 GeV energy range successfully
\cite{abdalla_2018_vela}. This
is the first time, measurements are reported from ground-based
atmospheric Cherenkov telescopes for this object.

\subsubsection{MAGIC}
\begin{figure}[h]
    \centering
    \includegraphics[width=0.65\textwidth]{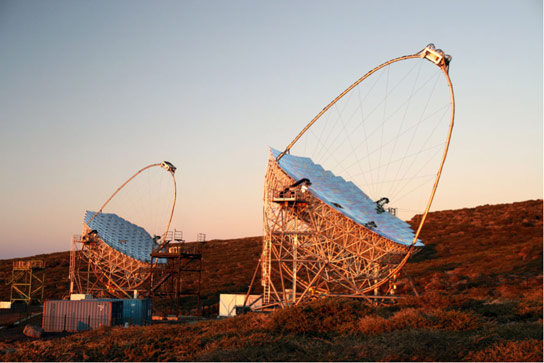}
    \caption{MAGIC telescope system at La Palma in the Canary Islands consisting of two 17m diameter telescopes. Picture courtesy : MAGIC collaboration.}
    \label{fig:magic}
\end{figure}

The Major Atmospheric Gamma Imaging Cherenkov \footnote{https://wwwmagic.mppmu.mpg.de} (MAGIC) telescope  is a system of two 17 m diameter IACTs (see figure \ref{fig:magic}) located in El Roque
de los Muchachos observatory in Canary Islands of La Palma
at an altitude of 2200 m a.s.l. 

These telescopes are designed in such a way that their lower parts, like 
undercarriage and bogeys which move only in the azimuth direction, are
made up of steel.
Whereas the lightweight upper parts such
as the mirror support are made up of carbon fibre reinforced plastic tubes. This unique
state-of-the-art design helps MAGIC telescopes to orient themselves by 180$^\circ$ in the azimuth
direction in about 25 seconds. MAGIC uses parabolic mirrors to reflect Cherenkov photons. 
The mirror surface is segmented in nature and is comprised of smaller units of 0.5-1$m^2$.

A single MAGIC camera comprises of 1039 PMTs of
Hamamatsu type R10408, 25.4 mm diameter and typical quantum efficiency of around
32\%. The PMTs or pixels are grouped in seven PMTs to build a modular unit. 
Each pixel has an FoV of 0.1$^\circ$ and the FoV of the MAGIC camera is 3.5$^\circ$.
(for details about the hardware of the MAGIC telescope system, see \cite{MAGIC_hardware}). 

Observations of the Crab Nebula were performed at low zenith angles to estimate the key performance 
parameters of the MAGIC stereo system. For these observations, the
standard trigger threshold of the MAGIC telescopes is $\sim$ 50 GeV. The integral sensitivity 
for point-like sources with Crab Nebula-like spectrum above 220 GeV is (0.66 $\pm$0.03)\% of 
Crab Nebula flux in 50 hours of observations \cite{MAGIC_software}.

The large size of MAGIC and its very fast repositioning scheme enables it to focus on 
observations of distant AGNs and GRBs as well as pulsars. MAGIC is the first TeV telescope
to have detected pulsations from the Crab with a high significance at energies greater 
than 25 GeV \cite{MAGIC_Crabpulsar1,MAGIC_Crabpulsar2}. 
Other significant detections include the first ever detection of TeV $\gamma$-ray 
photons from GRB 190114C\cite{MAGIC_grb} by an IACT, detection of a burst of photons at TeV energies from 
a candidate neutrino blazar (TXS0506+056)\cite{MAGIC_txs} and detection of high energy $\gamma$-rays 
from two of the most distant AGNs known, PKS1441+25\cite{MAGIC_pks1441} and QSO B0218+357\cite{MAGIC_b0218}. 
At the highest energies, taking advantage of very large zenith angle observations, MAGIC has been able to detect
$\gamma$-rays upto 100 TeV from the Crab nebula\cite{MAGIC_vlza}.

\subsubsection{VERITAS}
\begin{figure}[h]
    \centering
    \includegraphics[width=0.75\textwidth]{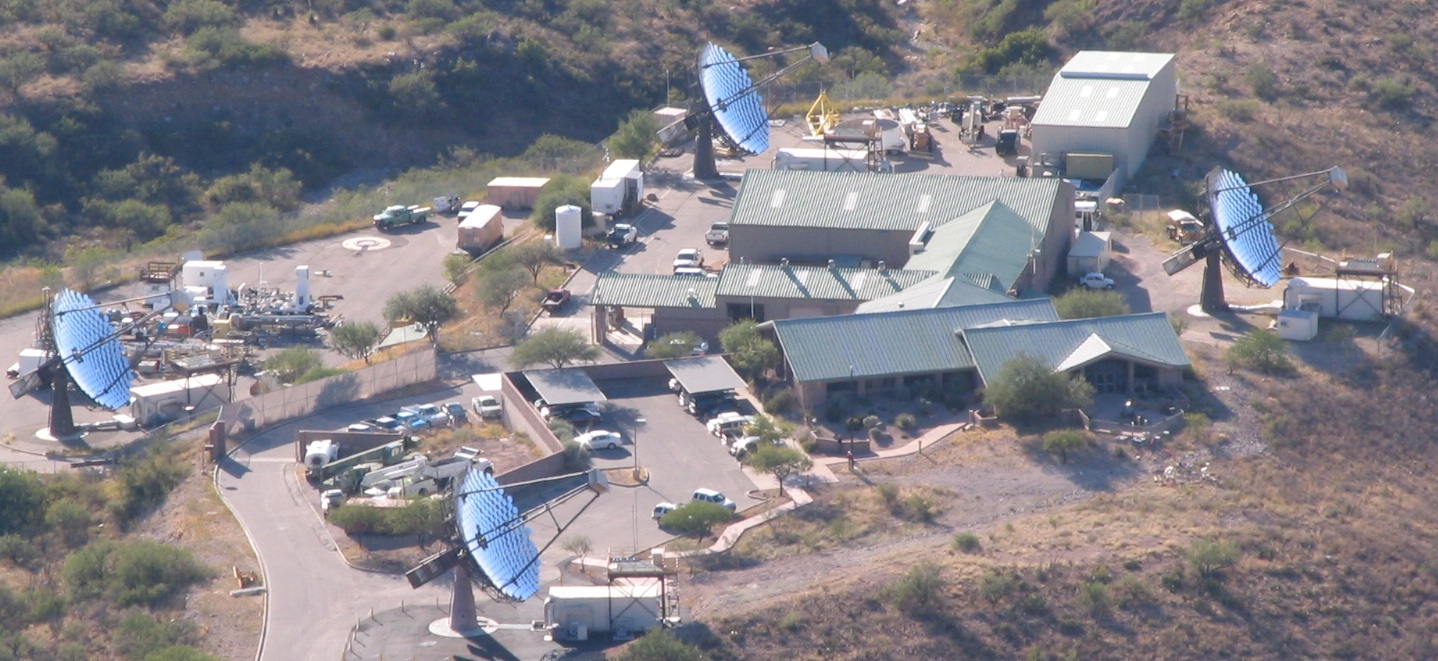}
    \caption{VERITAS array in Arizona consisting of four 12m diameter telescopes. Picture courtesy : VERITAS collaboration.}
    \label{fig:veritas}
\end{figure}

The Very Energetic Radiation Imaging Telescope Array System\footnote{https://veritas.sao.arizona.edu} (VERITAS)
is a ground-based 
$\gamma$-ray telescope array operating at Fred Lawrence Whipple Observatory (FLWO) 
in southern Arizona, USA. 
It is an array of four 12 m diameter reflectors (see figure \ref{fig:veritas}) designed 
to perform $\gamma$-ray astronomy of celestial sources in the energy range of approximately
100 GeV to several tens of TeV. 
Each telescope of VERITAS is based on an alt-azimuth positioner system with a maximum 
slew speed of about 1 degree per second. The pointing accuracy of a VERITAS telescope 
is typically better than $\pm$0.01$^\circ$. The optical system of the telescope is based 
on Davies Cotton design and consists of about 350 identical hexagonal spherical mirrors 
of 30 $cm^2$ each having a radius of curvature of about 24 m. 
Each mirror facet is made from glass, slumped, polished and then aluminized and anodized. 

The camera of the telescope is located at the focal plane of the telescope and has a 
dimension of about 1.8m $\times$ 1.8m. The camera is equipped with 499 circular PMTs, giving an angular pixel spacing of 0.15$^\circ$ and a total FoV of 3.5$^\circ$. 

VERITAS employs a three-tier trigger system to reduce the rate of several background events. 
The first trigger system works on the single-pixel level, the second checks for 
specific patterns of single-level pixels within a certain time window, and the 
third works as regards to telescope coincidence, requiring simultaneous observations 
of an air-shower event in at least two of the four telescopes (ensuring a "stereoscopic" 
view of the event). The angular resolution of VERITAS (68\% containment radius) is 
0.08$^\circ$ at 1 TeV and it worsens to about 0.13$^\circ$ at 200 GeV. The energy resolution is 
about 17\% at 1 TeV and the point source sensitivity is about 1\% of the Crab nebula flux 
in less than 25 hours.  

VERITAS carried out the first high-resolution scan of the Cygnus region in the northern
hemisphere and discovered a few sources in the region \cite{VERITAS_gammaCygni,VERITAS_CTA1,VERITAS_j2019}. 
Apart from that, one of the main highlights of the VERITAS discovery program has been the detection of a starburst galaxy
M82 \cite{veritas_m82}. Besides these interesting discoveries, it has also discovered a few AGNs at moderate
to high redshifts \cite{VERITAS_rbs0413,VERITAS_1es0414,VERITAS_pks1441} and also the Crab 
pulsar \cite{VERITAS_Crabpulsar} at energies upto several hundreds of GeV.  

\subsubsection{Other IACTs}

The Collaboration between Australia and Nippon for a GAmma-Ray Observatory in the Outback (CANGAROO-III), a stereoscopic array of imaging telescopes, was  operated
at Woomera in Australia at an altitude of 165 m a.s.l. during 2004 - 2011 \cite{Enomoto}.
This was an array of four telescopes with a reflector diameter of 10 m. One
moderate size telescope, TeV Atmospheric Cherenkov Telescope with Imaging Camera
(TACTIC) is operational at Mt Abu, in Western
India since 2001 \cite{Koul}. It has a 4 m diameter reflector with 349-pixel camera.

Another large size telescope based on imaging technique, Major Atmospheric
Cherenkov Experiment (MACE), is recently commissioned at Hanle \cite{Borwankar} in the Himalayas, India. The diameter
of its reflector is 21 m and it has a camera consisting of 1088 pixels.
Science observations with MACE are  expected to commence soon.
Based on the simulations, MACE is expected to achieve an energy
threshold of $\sim$ 20 GeV, taking advantage of high altitude location as
well as the large size of the reflector.

The First g-Apd Cherenkov Telescope (FACT) \cite{Anderhub} is the first IACT using 
Geiger-mode avalanche photodiodes (G-APDs) or silicon photomultipliers (SiPMs)
as photo-sensors instead of usual PMTs. SiPMs have several advantages
over PMTs, like higher photon detection efficiency, lower operating
voltage etc. Most importantly, unlike PMTs, they can be safely operated
in a bright environment. So it is possible to increase the observation duty
cycle by operating a telescope in the moon-lit part of the night partially.
FACT is a moderate size telescope with about a 4 m diameter reflector.
In fact, it is one of the mounts from the HEGRA telescope array which is
refurbished. Its camera consists of 1440 pixels of SiPMs with  solid light
concentrators mounted in front, to take care of dead space between the pixels.
FACT is operational since 2012 and has observed several sources, largely
blazar class AGNs and provided long duration observations for these
sources. This has paved a way for the usage of SiPMs in future generation
telescopes.

\subsection{Future IACTs : CTA}

Cherenkov Telescope Array\footnote{https://www.cta-observatory.org/} (CTA) project is a next-generation array of IACTs aimed at making measurements of the $\gamma$-ray sky with unprecedented details over a wide energy range from 20 GeV to $\sim$ 300 TeV\cite{acharya2013}. 
CTA will consist of two IACT arrays,
one in the southern hemisphere which will primarily focus on galactic $\gamma$-ray sources and the other in the northern hemisphere which will focus more 
on extra-galactic sources.
The expected sensitivity of CTA will be at least one order of magnitude better than any existing IACTs. CTA with its improved angular \& energy resolutions and wide field of view will study $\gamma$-ray sources with high precision. 

These arrays would allow for both deep field investigations and surveys in parallel with monitoring of the brightest
known variable sources and activity related to alerts from space-based instruments in other wavelengths. 
The site in the southern hemisphere is Paranal, Chile and the site in the northern hemisphere is La Palma, 
Canary Islands, Spain.
The two sites will allow for a full sky-coverage and consequently access to many more $\gamma$-ray sources
with the possibility of discovering many rare source classes and rare events, such as GRBs or supernova
explosions\cite{acharya2019}.

To reach the performance goals as mentioned above and specifically the wide energy range to be covered,
the design of the telescope array is required to be optimised for three adjacent energy ranges :

The low energy range $\leq$ 100 GeV : In order to detect $\gamma$-rays down to a few tens of GeV,
the Cherenkov light needs to be sampled efficiently as the various discriminating parameters between
a $\gamma$-ray induced shower and a hadron induced shower start to overlap at energies below 100 GeV. This creates a huge
background in the data and thus systematic uncertainties of the background estimation limit the
sensitivity of the instrument. This problem can be overcome with the deployment of a few closely
packed large-size telescopes to collect as many photons as possible from the low energy showers.

The core energy range 0.1 to 50 TeV : Shower detection and reconstruction in this energy range
has been the bread and butter of the current instruments. The sensitivity in this energy range can be
further improved by the deployment of an array of mid-size telescopes of about 12 m diameters with a
spacing of about 100 m. The array size for the first time will cover the entire Cherenkov pool which
will improve the sensitivity due to the increased collection area of the array and also due to higher quality
reconstruction of the shower parameters since individual showers will typically be stereoscopically
imaged by a large number of telescopes compared to a single large telescope or a few medium size
telescopes.

The high energy range $>$ 50 TeV : At the highest energies, the main limitation is the number of detected $\gamma$-ray showers due to the steeply falling flux. 
Hence there is a need for covering an area of several square kms. Deployment of an array of many small telescopes ($\sim$ 4 to 6 m class) over a large
area will extend the energy range beyond 300 TeV, a range where most imaging Cherenkov telescopes have never been operated.

It is clear from the current the knowledge of operation of Cherenkov telescopes that a mixed array is required
to achieve the best sensitivity over a wide energy range spanning over 4 decades in energy. Figure \ref{fig:cta}, shows the artistic illustration of the proposed CTA. The main design
drivers to cover the energy ranges mentioned above are the following :

\begin{figure}[h]
    \centering
    \includegraphics[width=0.7\textwidth]{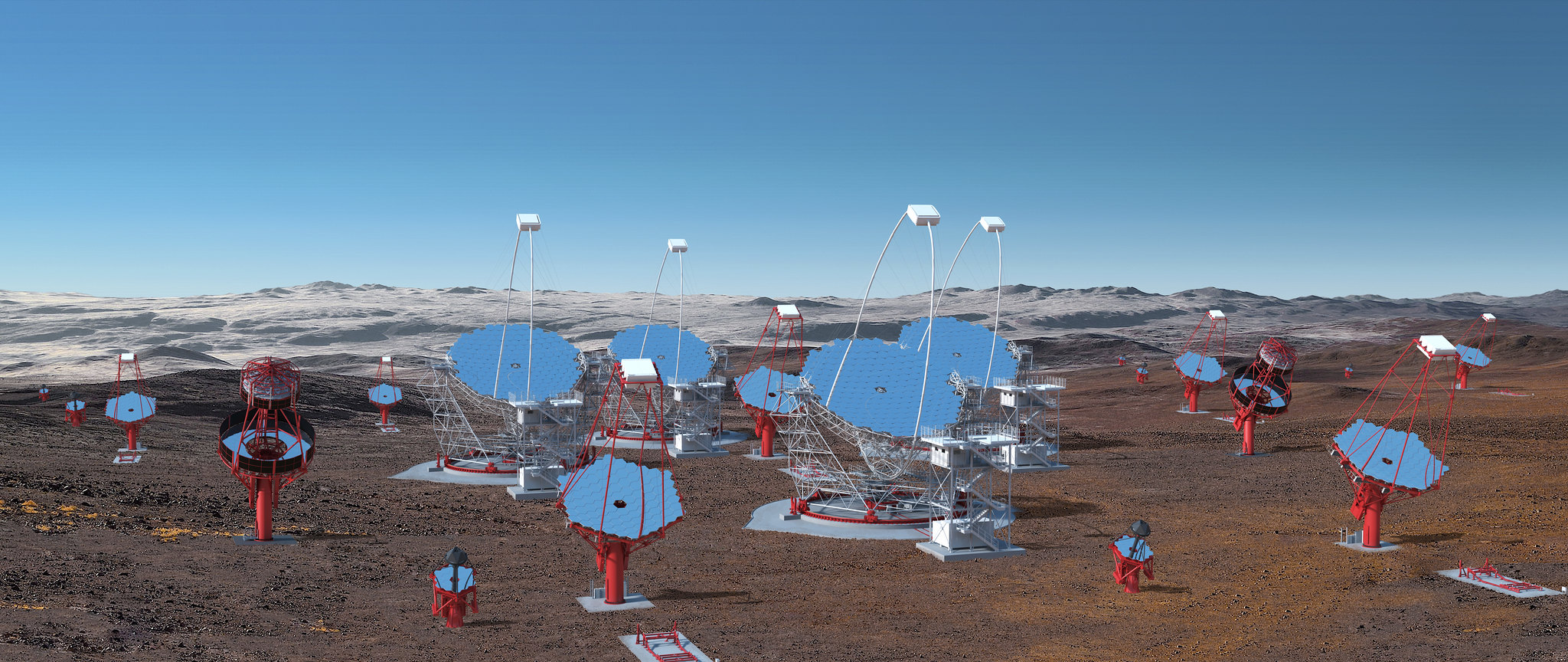}
    \caption{An artistic illustration of the proposed CTA. Image credit: Gabriel Pérez Diaz, IAC / Marc-André Besel, CTAO.}
    \label{fig:cta}
\end{figure}

\begin{figure}[h]
    \centering
    \includegraphics[width=0.7\textwidth]{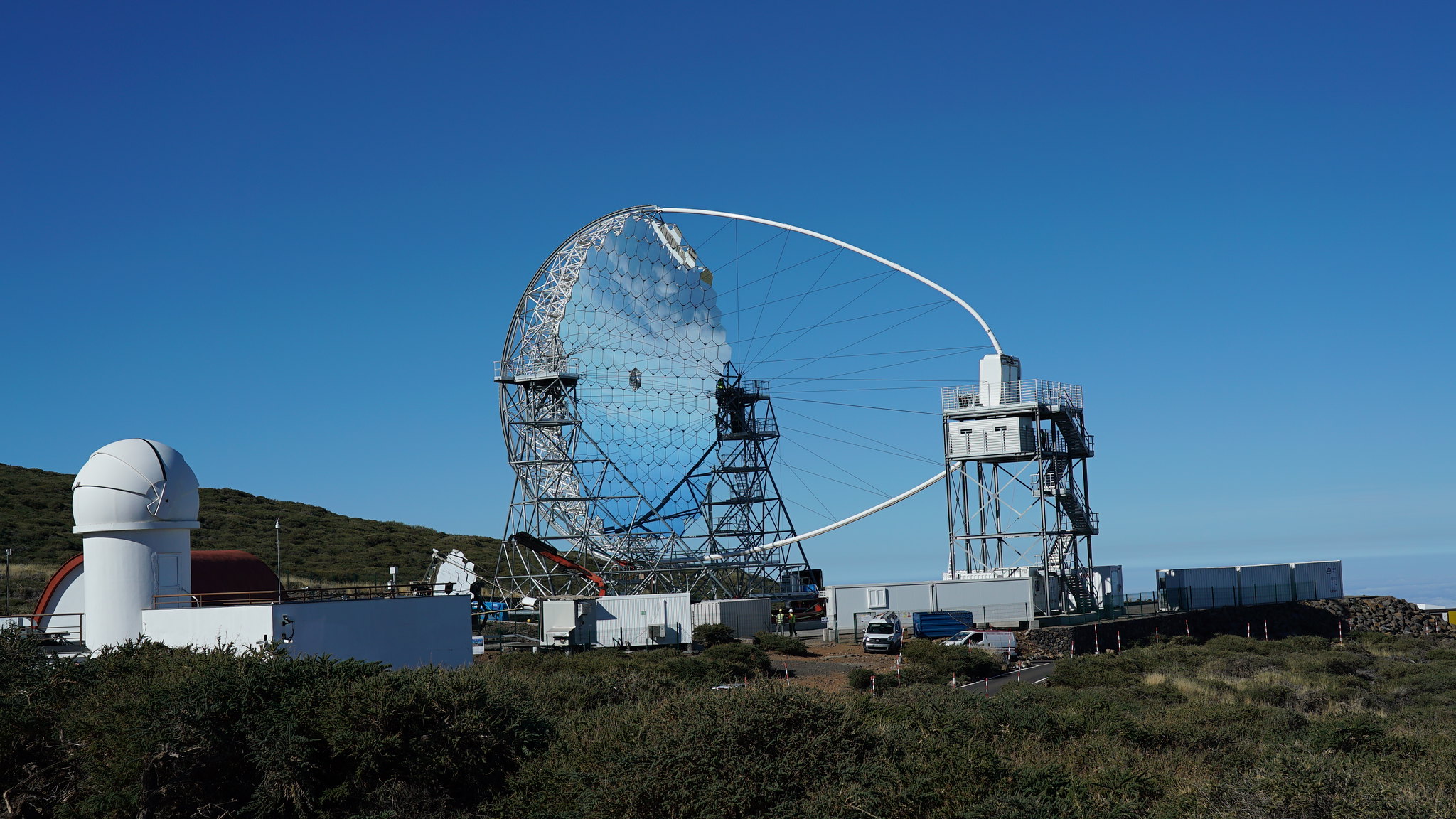}
    \caption{The proto-type Large Size Telescope (LST) installed in La Palma, Canary Island, Spain. Picture credit: Iván Jiménez (IAC), CTAO.}
    \label{fig:lst}
\end{figure}

Large Size Telescopes (LST) : A few large telescopes (23 m diameter class) with relatively narrow
FoV \footnote{The first LST constructed at La Palma has an FoV of 4.3 degrees} and fine pixelization ($\sim$ 0.1 degrees) 
with close spacing will enable one to access
the lowest energies possible and thus allow for detection of AGNs at high redshifts and galactic sources
like pulsars which exhibit strong spectral cut-offs at lower energies (few tens of GeV typically). Further,
at these energies, we will have a good overlap with satellite-based $\gamma$-ray detectors like Fermi-LAT.
The construction of the prototype LST\cite{cortina2019} is complete and it has been installed at La Palma, Canary Islands, Spain, 
on the CTA observatory site of the Observatorio del Roque de los Muchachos. It is currently being commissioned
and is foreseen to become the first CTA telescope at the site (see figure ~\ref{fig:lst}). 

Medium Size Telescopes (MST)
: In the known energy range between 100 GeV to 50 TeV,
where most of the current installations are operating, an extended array covering an area of few
$km^2$ is required for increased collection area providing a higher proportion of events which would be
contained within the array and would thus have excellent energy and angular resolution leading to
increased sensitivity. An array with few tens of telescopes of medium sizes (12 m diameter class) with
wide FoV (6 to 8 degrees) for studying morphologies of extended objects and moderate pixel size of 0.18 
degrees in under construction. Currently, a prototype MST has been built and several key parameters
of the telescope are being studied in detail. The camera of the prototype consists of about 1800
pixels with an FoV of 7.7 deg. The readout system is based on the philosophy called {\it FlashCam } 
where the trigger and readout system are fully digital based on FADCs and FPGAs as key components of 
front-end electronics\cite{flashcam2015}. An alternative camera based on the philosophy of {\it NectarCam}
is also under development. The NectarCAM camera has a modular design and each module consists of 7 PMTs associated with their readout and trigger electronics.  
Each module consists of a focal plane module, a front-end electronics board and a back-plane. The front-end board contains the NECTAr chips which perform 
the readout together with the analogue to digital conversion  and the local trigger electronics.\cite{nectarcam2019}

Small Size Telescopes (SST) : Detection at the highest energies from 50 TeV to 300 TeV encounters
the problem of the sparseness of the signal from extremely bright $\gamma$-ray sources. This leads to the
requirement of a few tens of telescopes of small sizes (5 m diameter class) covering an area of $\sim$
10 $km^2$ with very wide FoV (7 to 10 degrees) so that air showers can be seen in stereo by widely
separated detectors and relatively coarse pixelization (0.25 degrees approximately). Such an array is
mainly preferred for the southern observatory where galactic $\gamma$-ray sources are our prime targets.
The need for a large FoV due to large inter-telescope spacing would actually require a novel solution
for these types of telescopes which have been explored. The possibility of the use of a dual-mirror
Schwarzchild-Couder optical design with a very small plate scale thus allowing for a smaller camera
with novel technologies like Silicon photomultipliers is being explored\cite{asano2018}.

CTA will become
a worldwide project for the future VHE $\gamma$-ray astronomy with Cherenkov telescopes.
As the first phase of the project, CTA consortium has finalised the ``Alpha Configuration" \footnote{https://www.cta-observatory.org/science/ctao-performance/}, which will consist of 4 LSTs and 9 MSTs in the northern 
hemisphere and 14 MSTs and 37 SSTs in the southern hemisphere. 
CTA will not only
consolidate VHE $\gamma-$ray astronomy as a major branch of astronomy, but will be eventually considered as
one of the leading astronomical observatories of the world with tremendous synergies with other facilities in
other wavelengths, for example, ALMA (mm wavelength), Thirty Meter Telescope (TMT) in optical, Square
Kilometer Array (SKA) in radio, eROSITA in X-rays and Large Synoptic Survey Telescope (LSST) which
are expected to start operations in the next decade.


\section{Simulations and Data Analysis}
\label{sim}

Monte Carlo simulations of EAS form an integral part
of the atmospheric Cherenkov technique. In absence of direct calibrations,
simulations provide the only way to understand and estimate the performance
of the telescope system. Apart from this, simulations provide important
inputs for the design of the telescope system and play a major role in data
analysis and improvement of sensitivity of the instrument.

Several software packages were developed to simulate EAS  initiated by $\gamma$-rays as well as various cosmic ray species
and also to simulate Cherenkov light production by relativistic charged particles in
EAS. Some of the packages used for simulations are MOCCA \cite{mocca},
ALTAI \cite{altai}, KASCADE \cite{kascade} and CORSIKA \cite{corsika_1,corsika_2}. These were used extensively for simulations
of previous generation telescope systems. During the last fifteen years,
CORSIKA has emerged as a package used most extensively for air shower
simulations.

CORSIKA (i.e. Cosmic Ray Simulation for KAscade) is a detailed Monte
Carlo program which is used to study the evolution of EAS initiated by $\gamma$-rays and various particles like protons,
nuclei etc \cite{corsika_1,corsika_2} (see CORSIKA website\footnote{https://www.iap.kit.edu/corsika/}). Originally developed for simulations of the KASCADE
experiment at Karlsruhe, CORSIKA has evolved over the last three decades
since its first version in 1989 and by now it is used for simulations
of various atmospheric Cherenkov (E $\sim$ 10$^{12}$ eV),  air shower 
(E $\sim$ 10$^{20}$ eV) as well as neutrino experiments.

While simulating air showers initiated by various cosmic ray species,
CORSIKA deals with interactions between hadrons, nuclei, electrons,
photons etc. It uses a variety of models for low energy and high energy
hadronic interactions. For high energy hadronic interactions, one
can choose between VENUS, QGSJET, DPMJET based on Gribov-Regge theory,
SIBYLL which is mini-jet model and HDPM which is a phenomenological
generator adjusted with experimental data whenever possible, NEXUS
and EPOS based on accelerator data. Options available for low energy
hadronic interactions include GHEISHA which is a  Monte Carlo program,
FLUKA which is a refined model with many details of nuclear effects and
UrQMD which describes microscopically the low energetic hadron-nucleus
collisions. Further, CORSIKA uses EGS4 code for treating interactions
of electrons and photons. In addition to this, simulation of
Cherenkov photon production by relativistic particles in EAS is also available as a part of CORSIKA. IACT routines
provided by K. Bernlohr are particularly useful for treating
Cherenkov radiation for IACTs \cite{iact}.

\begin{figure}%
    \centering
    {{\includegraphics[width=3.5cm]{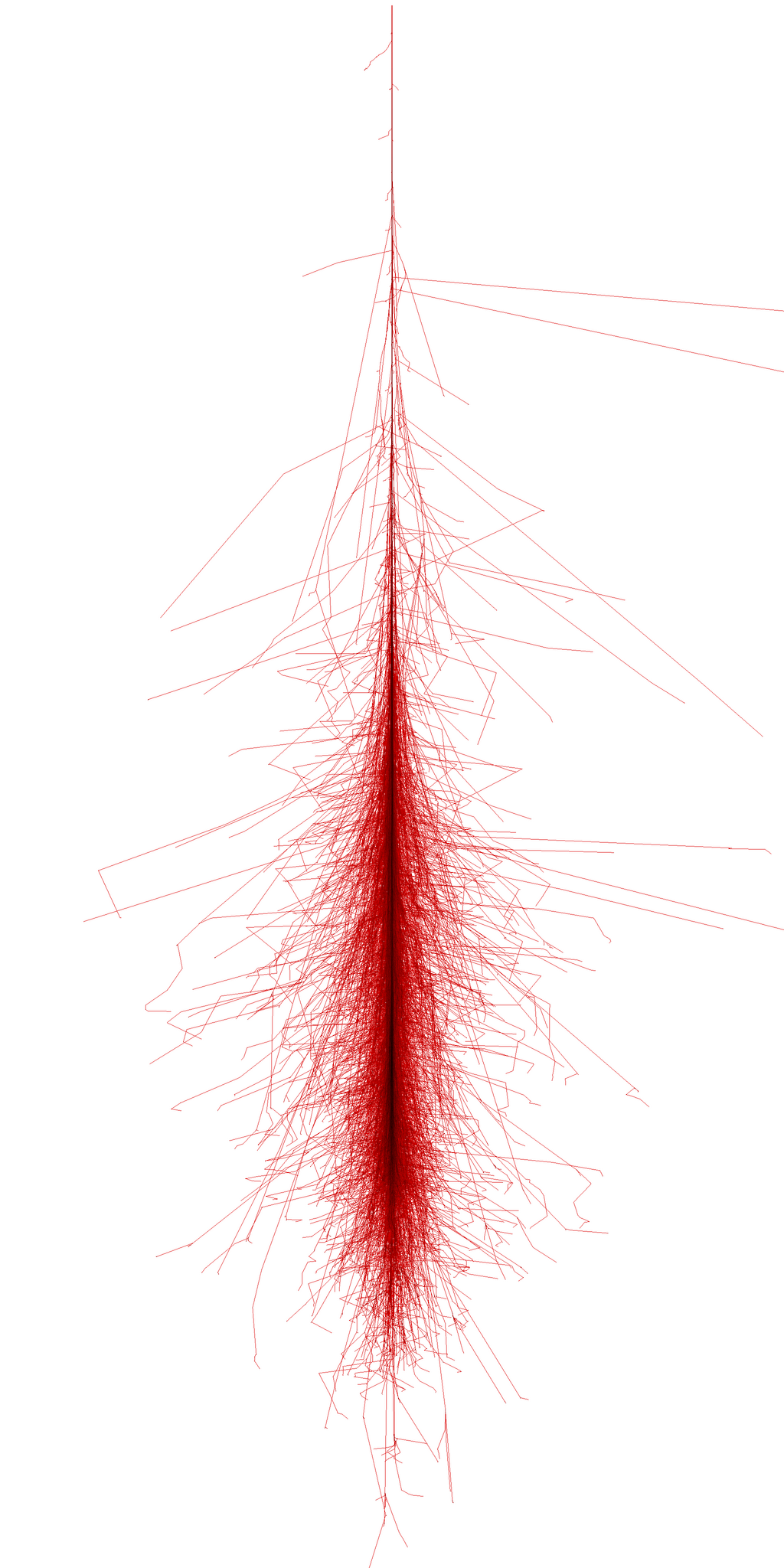} }}%
    \qquad
    {{\includegraphics[width=3.5cm]{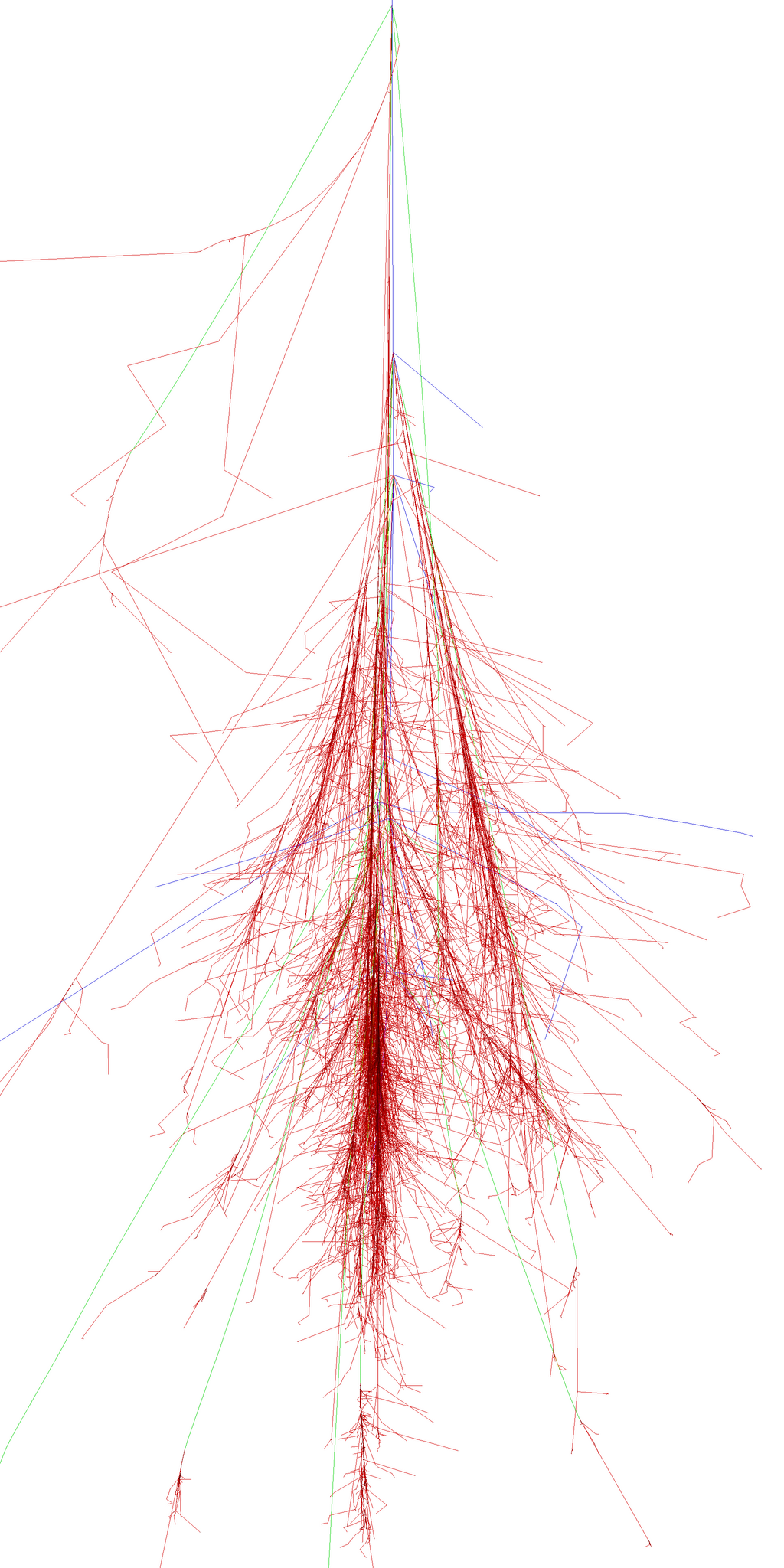} }}%
    \caption{Left : EAS generated by 100 GeV $\gamma$-ray, right : EAS generated
    by 100 GeV proton primary. Image Credit : CORSIKA website  https://www.iap.kit.edu/corsika}.
    \label{fig:simulation}%
\end{figure}

CORSIKA provides an option between various atmospheric profiles such as
US standard atmospheric profile parameterised by J. Linsley, profiles
for Central Europe, south Pole, tropical region etc. Atmospheric
density variation is modelled in these profiles which is crucial
for the evolution of shower and Cherenkov production. Depending on
observation location, the appropriate profile can be selected. CORSIKA
also takes into account the the deflection of charged particles in the shower by
geomagnetic field at the location of the telescope. Wavelength dependent
absorptions of Cherenkov photons such as atmospheric absorption,
mirror reflectivity, the quantum efficiency of PMT are also implemented
in the code. Finally, geometry of the telescope array can be given as
input parameters.

At the output, CORSIKA gives the distribution of particles, $\gamma$-ray
photons and Cherenkov photons reaching the  observation level. For each
Cherenkov photon hitting telescopes, its location, arrival time,
arrival direction and production height is recorded in the output file.

Figure \ref{fig:simulation} shows typical simulated cascades for $\gamma$-ray and proton primaries.
Combining Cherenkov photon distribution with detector specific
features such as pulse shape, trigger criteria, NSB
at observation site etc performance parameters 
are evaluated. GHS parameters, which play a major
role in improving sensitivity for detection of $\gamma$-ray signal
against cosmic ray background are also obtained from simulations.

Performance parameters indicate various characteristics of the 
telescope system. These parameters are used in the analysis of the data on celestial objects recorded by telescopes and are also useful for comparing various telescope systems. These parameters, obtained from
simulations, are described below.

The first parameter is energy threshold i.e. the lowest energy
of $\gamma$-ray  that can be detected by the telescope. It depends on several parameters specific to the telescope system such as the size of the reflector, quantum efficiency of PMTs, NSB at observation site, trigger criteria, observation site altitude etc. Typically peak of the differential rate curve generated with simulated $\gamma$-ray showers is used as a measure of energy threshold. 

The second parameter is the effective collection area, mentioned 
earlier. This area is much larger than the physical area of 
the light collector or reflector
and is comparable to the size of the Cherenkov pool, which roughly
corresponds to a circular area with a radius of about 120 m at
observatory locations at moderate altitudes. The effective collection 
area is energy dependent, largely decided by the trigger efficiency. 
It is given by the following equation,

\begin{equation}\label{eq:1}
A_{eff}(E) = 2\pi \int_{0}^{\infty} R \eta(R,E) dR
\end{equation}
where, $\eta(R,E)$ is the trigger probability, R is the distance from shower core. 

The next parameter is sensitivity for detection of $\gamma$-ray signal 
in presence of background produced by cosmic rays. Apart from
telescope parameters, it largely depends on the ability to reject 
cosmic ray background while analysing the data, achieved using
various GHS parameters. 

The sensitivity of the instrument is defined as the minimum flux from a steady point-like 
source to be detected in 50 hours for 5 standard deviations and is estimated in narrow bins of energy.
When the background is perfectly well estimated, the significance of the detection of a source can be computed
using a simple formula: 
\begin{equation}\label{eq:2}
S = \frac{N_{\gamma}}{\sqrt{N_{cr}}} 
\end{equation}
where $N_{\gamma}$ and $N_{cr}$ are the excess and background events respectively. One can thus define the 
sensitivity S as the flux of the source yielding ${N_{\gamma}}/{\sqrt{N_{cr}}}$ = 5 after 50 hours of observation.
However, the sensitivity is more realistically calculated using Equation 17 of \cite{LiMa1983} which is a
standard method in calculating significances in VHE $\gamma$-ray astronomy. To assess the performance of an IACT for
sources with any shape, one determines the differential sensitivity calculated in narrow bins of energy (typically
4 or 5 bins per decade). Often several other conditions like ${N_{\gamma}} > 10$ and  ${N_{\gamma}} > 0.05 N_{cr}$ 
are imposed during the estimation of the sensitivity. The first condition ensures that we are dealing with Gaussian 
distribution and the second condition ensures that one is not limited by the systematics of background determination
which may happen when a small signal resides over a large background. 

Simulations play a major role in the analysis of data from wavefront
sampling arrays as well as IACTs. Apart from providing performance
parameters, they have played an important role in improving the sensitivity of IACTs.
In the next few paragraphs, we outline the procedure for the analysis of data
from IACTs.

The raw data in the case of an IACT consists of a digitally sampled trace of a signal for each 
PMT in the camera. In the first step, the pedestal (a baseline value
in the absence of any Cherenkov photon) is subtracted during the extraction of the signal. 
In the next step, those pixels which contain information about Cherenkov photons above a 
pre-defined threshold are identified. The Cherenkov photons produced by the EAS trigger pixels and thus an image of the shower is formed along with 
NSB and electronic noise. Hence the image from the shower needs to be cleaned. The resulting
cleaned image is then parameterized based on a moment fitting approach, commonly known
as "Hillas parametrization", briefly mentioned in the previous section. It is the most commonly used method given the elliptical nature 
of the shower images, as shown in figure \ref{fig:imaging}. An alternative technique has also been later developed based on semi-analytical 
method of fitting shower images\cite{lebohec1998,lemoine2006,naurois2009}. 
In the commonly used method of image parameterization, several parameters of the image are calculated (parameters are listed in table \ref{tab1}). 

\begin{table}
\begin{tabular}{|c|p{9cm}|}
\hline
\textbf{Parameters}&  \textbf{Description of the parameters }  \\
\hline
Size     &       Total number of phes in the shower image. Size is proportional (at first
approximation) to the energy of $\gamma$-ray photon that induced the shower.\\ \hline 
CoG   &       Centre of gravity of the image formed at the camera. This parameter
determines the position of the weighted mean signal in the camera.  \\ \hline
Length   &  Semi-major axis of the elliptical fit to the image. This is related to the longitudinal development of the shower. \\ \hline
Width  &    Semi-minor axis of the elliptical fit to the image. This is related to the lateral development of the shower \\ \hline
Conc-n  &   Fraction of the phe-s contained in the n brightest pixels. \\ \hline
Leakage &  Fraction of signal in the outer rings of the image to the total size. only a part of the showers with large impact parameters are captured in the images leading to a wrong estimation in energy. Low leakage value means the image can be properly paramterized and energy determined properly. \\ \hline
Alpha & The angle between the major-axis and the line joining the CoG with the position of the source in the camera \\ \hline
Dist & The distance between the CoG and the position of the source in the camera. \\ \hline
Time gradient & The arrival time of the events in each pixel along the major-axis is fitted with a linear function. The linear coefficient is termed as the time gradient\\ \hline
Time RMS & The spread of the arrival times of the pixels in the image \\ \hline
\end{tabular}
\caption{{\label{tab1}}Several image parameters used for the paramterization of the 
image on the camera plane. These parameters are later used in the analysis. }
\end{table}

For a single telescope, the shower images are parameterized where parameters of the 
shower like {\it length}, {\it width} and orientation of the shower are calculated \cite{Hillas}. The
cosmic ray showers which produce similar Cherenkov images on the camera plane far outnumber
the $\gamma$-ray showers. Using the above mentioned parameters, the showers induced by the two primaries can 
be well discriminated. The power of discrimination improves tremendously when two or more  telescopes are deployed 
and multiple telescopes view the 
same shower \cite{daum1997} in stereoscopic technique pioneered by HEGRA. The stereoscopic observation allows one to perform a purely  geometrical reconstruction of the shower in 3-dimensions. The images from multiple telescopes can suitably be combined to give very useful information regarding the shower core, shower axis direction and height of the shower maximum and hence the stereoscopic method vastly 
improves the sensitivity of the telescope system. 

Most of the events registered by an IACT are produced by cosmic rays and hence $\gamma$-ray signal
has to be extracted by suitably rejecting all the cosmic ray showers on the basis
of information from the image shape and reconstructed direction. Several methods are employed 
to perform GHS effectively. The simplest of them is to apply cuts judiciously
on length and width parameters. This was initially developed by the Whipple group when the observations
on putative $\gamma$-ray sources were carried out by a single telescope. For stereoscopic observations, 
the determination of the shower
core can be performed from the intersection of the major axes of the images. Once the shower core 
is reconstructed, the measured widths and lengths of the images can be compared with the 
width and length parameters generated by Monte Carlo simulations, for images with the 
same Cherenkov intensity. As a result of this comparison, a quantity called ``mean scaled width"
and ``mean scaled length" can be defined which are used to provide discrimination between
$\gamma$-ray induced and cosmic ray induced showers \cite{daum1997,hegra1999}.
Another very interesting method was first developed by the CAT collaboration where a model of the 
shower development based on the knowledge of primary energy, arrival direction and impact parameter
was used. A simple analytical two dimensional model of shower profile was successfully 
implemented \cite{lebohec1998}. This method
was further refined to work for multiple telescopes' three dimensional analytical model of shower development\cite{lemoine2006}. 
Other methods like a log-likelihood approach using information from all pixels in the camera\cite{naurois2009} and 
comparisons of template images using Monte Carlo simulations have also been used where it has been shown that
the sensitivities obtained from these methods may be better than the simple Hillas parameterization method. 
More advanced techniques have recently been 
developed by various groups based on multivariate analysis like Random Forest, Neural Netowrk etc\cite{albert2008}.
Whatever be the method of discriminating $\gamma$-ray and cosmic ray induced showers, most of them 
have shown very high background rejection efficiency ($>$ 99\%) while retaining 75\%-80\% of the $\gamma$-rays. 
Stronger cuts with lower $\gamma$-ray efficiencies are also quite often used for discovering a weak 
source. 

\begin{figure}[h]
    \centering
    \includegraphics[width=0.65\textwidth]{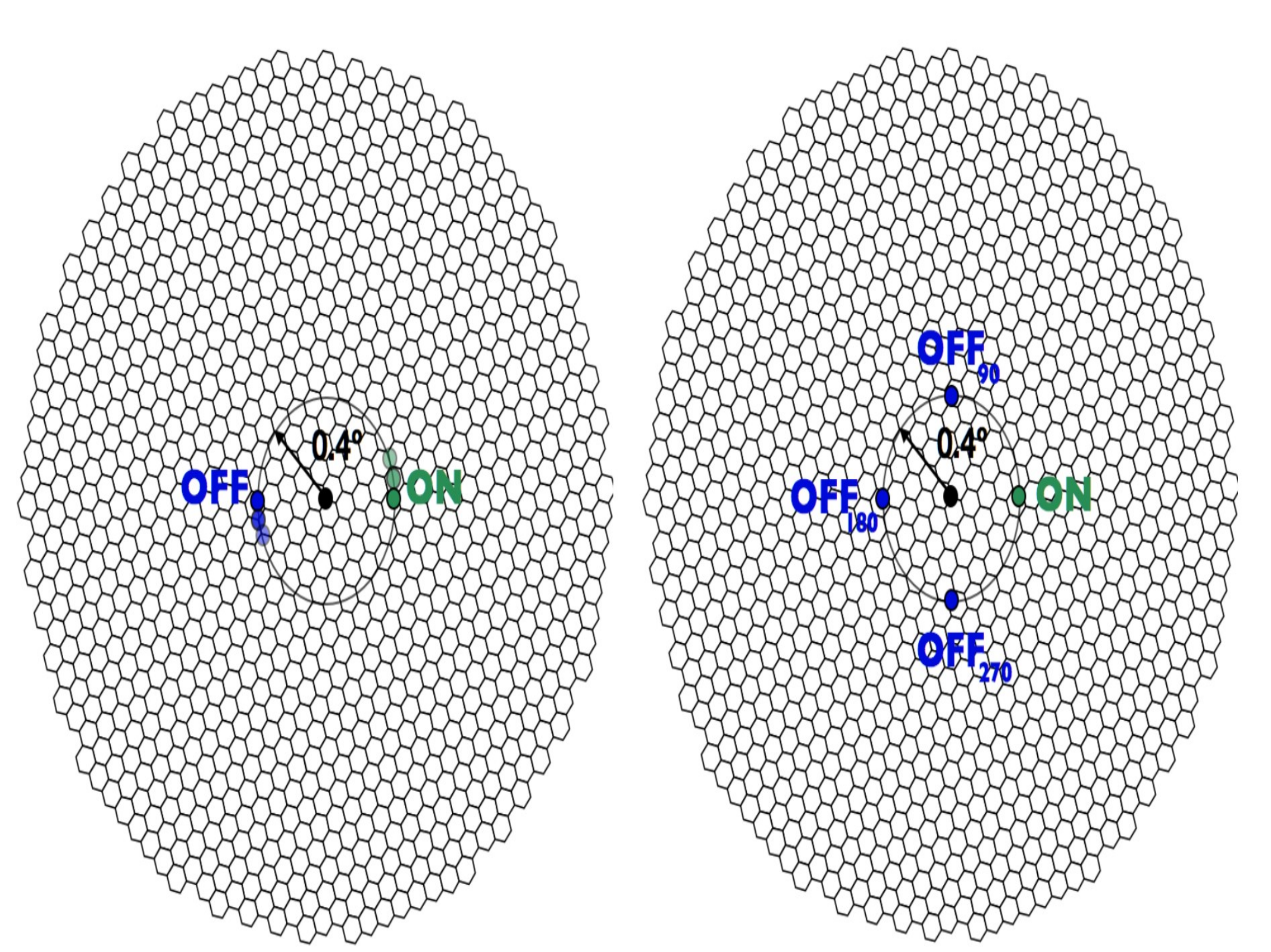}
    \caption{A typical picture of a wobble mode observation where the telescope is pointed with a slight offset from the source  and the background region [one (a) or many (b)] can be estimated simultaneously during ON-source observations. This figure is reproduced with permission from Dr. Ruben Lopez Coto's thesis.}
    \label{fig:wobble}
\end{figure}

The standard mode observation in IACT is the so called ``wobble-mode" \cite{fomin1994}. 
In this method, the telescopes are pointed slightly away from the source position and hence
allows the observations of ON and OFF regions simultaneously (see figure \ref{fig:wobble}).
Thus the systematics involved in the determination of the background can be controlled and it also 
allows for a more effective use of the observation time. The choice of this offset is optimized 
based on two effects: 1) a too small offset may result in overlap between the ON and 
OFF region which degrades background estimation, and 2) a too large offset may strongly affect 
the detection efficiency of the source.
Depending on the extension of the source region, multiple
OFF regions can be selected. For a single OFF position, the diametrically opposite 
position with respect to the camera center can be selected. 
Multiple OFF positions help in the estimation of the background more precisely. These
positions are selected symmetrically around the source position in such a way that any
inhomogeneity in the field of view of the camera is minimized.

\begin{figure}[h]
    \centering
    \includegraphics[width=0.65\textwidth]{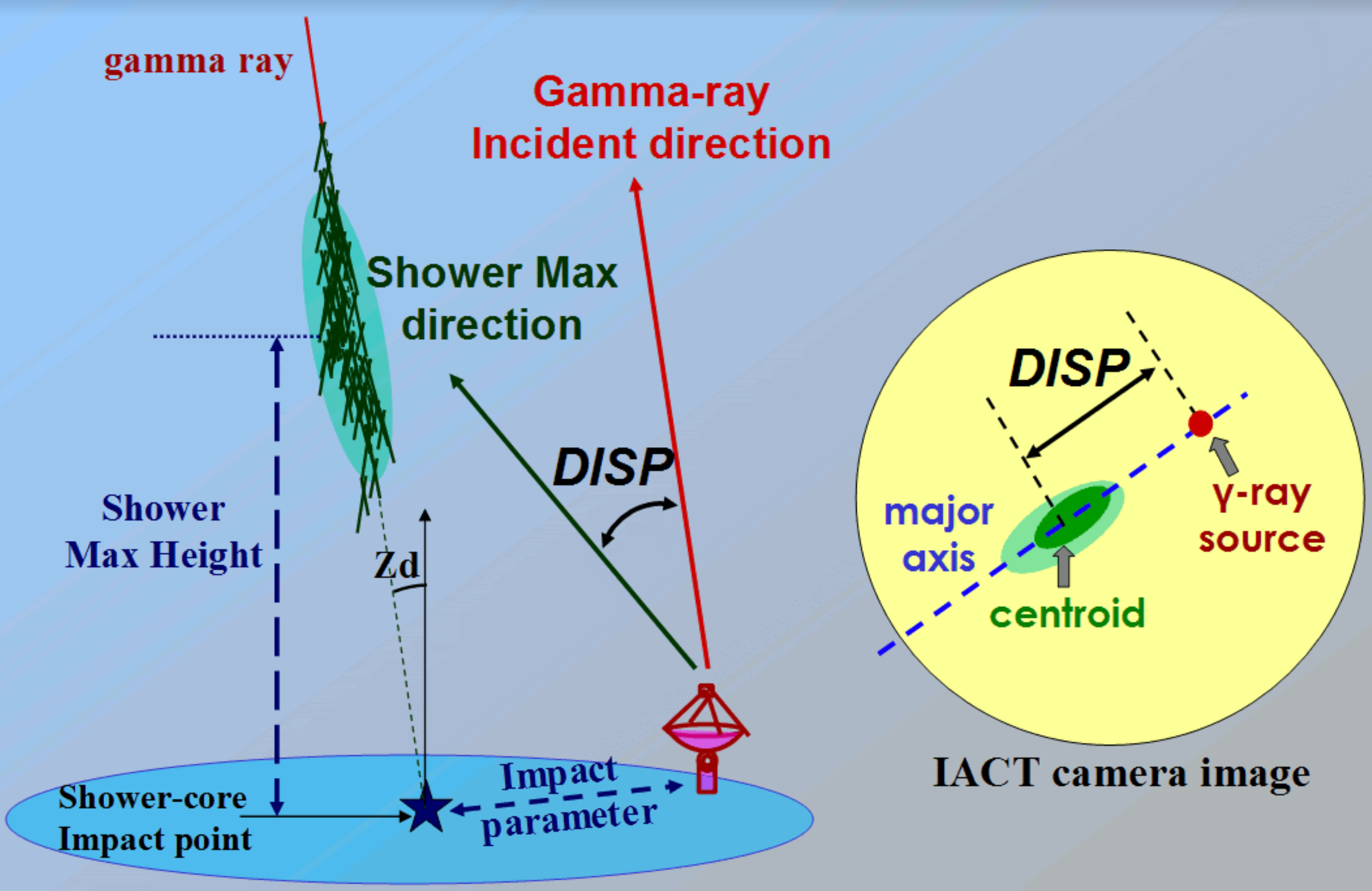}
    \caption{A schematic diagram of the DISP (Distance between the Image centroid and the Source Position) parameter. 
    The main principle pf the method is to calculate the DISP parameters and then determine the position of the incident 
    $\gamma$-ray direction on the main image axis at the distance DISP from the image centroid. The authors acknowledge the MAGIC collaboration for providing this diagram.}
    \label{fig:disp}
\end{figure}

The reconstruction of the arrival direction of the shower enables one to discriminate between
$\gamma$-ray photons from the putative source from the abundant isotropic charged cosmic rays.
It also helps one to map out the $\gamma$-ray emission in case of extended astrophysical sources,
especially in our Galaxy which can provide valuable clues to the particle acceleration mechanisms
in the source. The angular resolution is energy dependent and one can attain a very good
resolution of better than 0.1$^\circ$ for 68\% of the $\gamma$-rays from a point source at 1 TeV.
At lower energies, the angular resolution tends to degrade owing to the lower number of
Cherenkov photons available and due to shower to shower fluctuations.

The event-wise direction reconstruction is estimated using the so-called DISP (Distance between the Image centroid and the Source Position) method, 
where DISP parameter is defined as the distance between the CoG (Centre of Gravity) and the impact point (see Fig.~\ref{fig:disp}). 
Using this method, the position of the incident $\gamma$-ray direction on the main image axis 
at the distance from the image centroid is determined. In this case, two positions
on each side of the image centroid are possible. Further, the head-tail discrimination is
performed using the image parameters that characterize the asymmetry of the image. 
Typically, the time gradient of the development of the shower along the major axis or the 
third moment is used to perform this discrimination and hence determine the most probable 
$\gamma$-ray direction. 

The angular resolution is typically defined as the angular distance around the source which
contains 68\% of the excess $\gamma$-ray events. However, in several cases, it is also defined
as the standard deviation of a 2-dimensional Gaussian fitted to the distribution of reconstructed
excess events. Such a 2-D Gaussian corresponds to an exponential fit for the distribution
and corresponds to approximately 39\% containment of the excess events.

The energy of the event is reconstructed using suitable Look-up Tables (LUTs), 
a multi-dimensional table containing mean energy for each combination of the image parameters.
The LUTs are created using simulated $\gamma$-rays relating the event
energy to the impact and Cherenkov photon density measured by each telescope. Corrections
due to zenith, azimuth and large images which are partially contained in the camera are 
applied. The final estimated energy is computed as
the average of the energies reconstructed individually for each
telescope, weighted by the inverse of their uncertainties.
Typical energy resolutions of current generation IACTs are about 15\% at 1 TeV and 
degrade to around 25\% at lower energies ($\sim$ 100 GeV).  

\begin{figure}[h]
    \centering
    \includegraphics[width=0.65\textwidth]{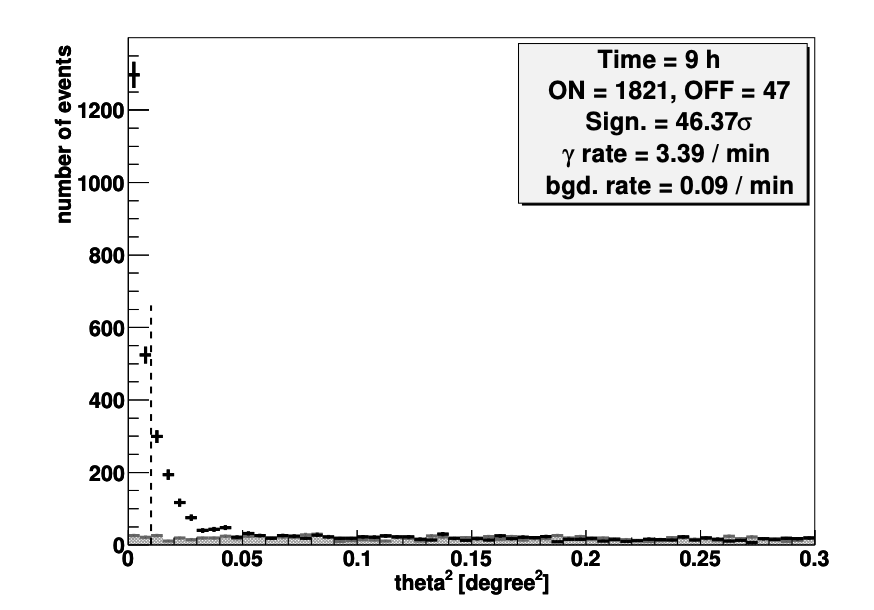}
    \caption{A typical $\theta^2$ distribution from the direction of the Crab nebula showing the signal events (black) and the background events (grey)
    above an estimated energy of 300 GeV. The vertical dashed line shows the cut on $\theta^2$ applied to calculate the significance from
    signal and background region. The figure is reproduced with permission from \cite{magic_thetasq} }
    \label{fig:thetasq}
\end{figure}

In order to estimate the excess events from a putative source and also calculate its significance,
the squared angular distance between the nominal source position in the camera and the
reconstructed source position is used. This quantity, called $\theta^2$, is then plotted. 
The background is estimated by potting the $\theta^2$ distribution in the OFF-regions, as shown in the figure \ref{fig:thetasq}. After
suitable normalisation is done in the region where one expects no signal or excess, the 
number of excess events are then calculated which is the difference between the number
of ON events and the number of OFF events. The significance of the excess is calculated using
Li and Ma\cite{LiMa1983}. 

In order to estimate the flux of $\gamma$-rays from a source, the knowledge of effective 
area of the detector is required. 
The effective area is energy dependent and is calculated using simulated $\gamma$-ray showers
over a wide range of impact parameters and an energy distribution close to that of the source
(typically a power law at TeV energies). 

\begin{figure}[h]
    \centering
    \includegraphics[width=0.95\textwidth]{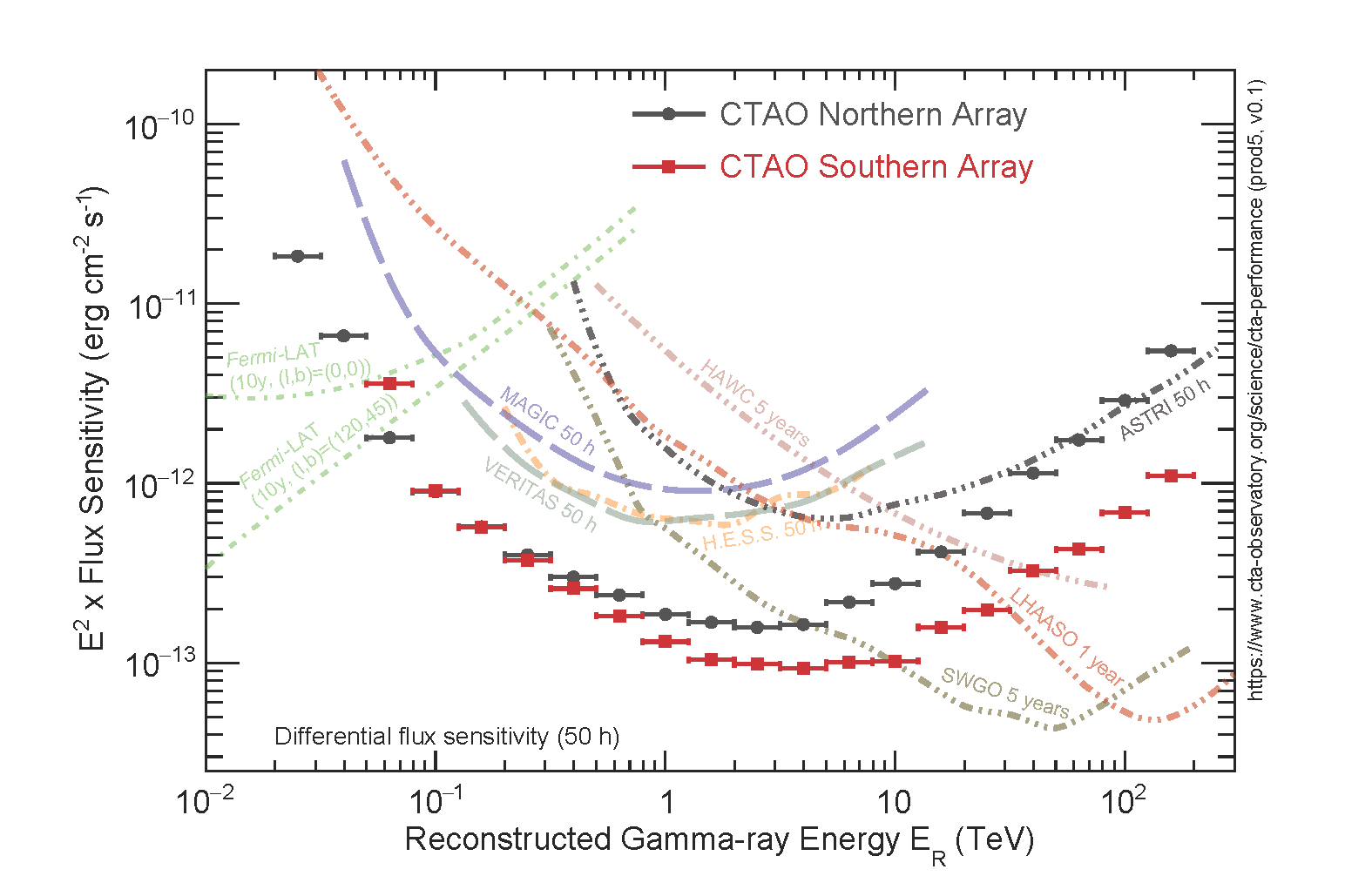}
    \caption{Differential flux sensitivity of the ``Alpha Configuration" of CTA arrays. The sensitivities of the current generation
    of IACT, high energy $\gamma$-ray instruments like Fermi-LAT and very high energy $\gamma$-ray instruments like HAWC are also shown for
    comparsions. The figure is reproduced with permission from CTA collaboration (https://www.cta-observatory.org/science/ctao-performance/)}
    \label{fig:CTA_sensitivity}
\end{figure}

 As has been explained earlier, the sensitivity of a VHE $\gamma-$ray instrument is
defined as the minimum flux of $\gamma-$rays observed from a putative source at a significance
of 5$\sigma$ in a typical observation time of 50 hours. Figure~\ref{fig:CTA_sensitivity} shows the
differential flux sensitivities (sensitivity in bins of energy) of the 
``Alpha Configuration" of CTA arrays w.r.t. the 
present generation of 
IACTs.
Along with that, we have also plotted the sensitivities of space-based detectors (for eg. Fermi-LAT)
and ground-based air shower array detectors (for eg. HAWC) for various time scales.
Typically a satellite-based detector has a lower $\gamma-$ray collection efficiency owing to the
limited size of the detector and hence requires longer integration times in order to
detect an appreciable number of high energy photons. This limits the sensitivity of
the space-based detectors at high energies beyond a few GeV. The sensitivities of the
air shower array detectors are inferior to the VHE $\gamma-$ray detectors (see figure~\ref{fig:sensitivity}) owing to the fact that
they have worse gamma-hadron separation and poorer angular resolution and hence also require
longer observation times (typically about a year or more) in order to detect a flux comparable to
the IACTs. The sensitivities of the IACTs start to worsen at energies around a few tens of GeV 
owing to the poor gamma-hadron separation at these energies and also at energies beyond a few tens
of TeV because at these energies the sensitivity is limited by the statistics 
of the number of photons from the source. The best operating range of the IACTs is from 50 GeV
to 50 TeV where most of the important discoveries in the field have been made.

\section{Air shower arrays}
\label{eas}
Apart from the detection of Cherenkov light, there is a complementary way to detect $\gamma$-rays on the ground indirectly, by detecting charged secondaries present in the EAS. These secondaries can be detected in two ways:  Water Cherenkov Technique (WCT) and Extensive Air Shower (EAS) array technique. In WCT, detectors  located inside the water tank or pool detect Cherenkov radiation produced by charged particles present in the shower. For the EAS array technique, charged particles are detected by an array of particle detectors like  scintillation detectors spread over a large area. Both these techniques require secondary charged particles at the observation level for a trigger. Therefore they have a higher energy threshold compared to ACTs and are usually located at very high altitudes. 
The existence of a large number of muons in hadronic air showers compared to $\gamma$-ray showers provides air shower arrays with a tool to distinguish between these two types of primaries.
For an EAS initiated by a vertically incident $\gamma$-ray with energy 100 TeV, approximately 20,000 shower particles will survive at an altitude of 2 km and for a 1 TeV $\gamma$-ray there will be a few 
hundreds of charged particles only. The corresponding numbers at an altitude of 4.5 km would be 100,000 for a 100 TeV $\gamma$-ray and a couple of thousands for a 1 TeV $\gamma$-ray induced showers. Even though the energy thresholds of air shower experiments are higher, they have certain advantages over ACTs. These detectors can operate $24\times7$ and their field of view is 2 sr i.e. they can observe half of the sky. Angular resolutions of air shower arrays are inferior to ACTs. However, because of large sky coverage and almost 100\% duty cycle, they are useful for the detection of  transient sources and sky surveys. 
These arrays are particularly suitable for searching PeVatrons in the Milky Way.

\subsection{Water Cherenkov Technique }
\begin{figure}[h]
    \centering
    \includegraphics[width=0.65\textwidth]{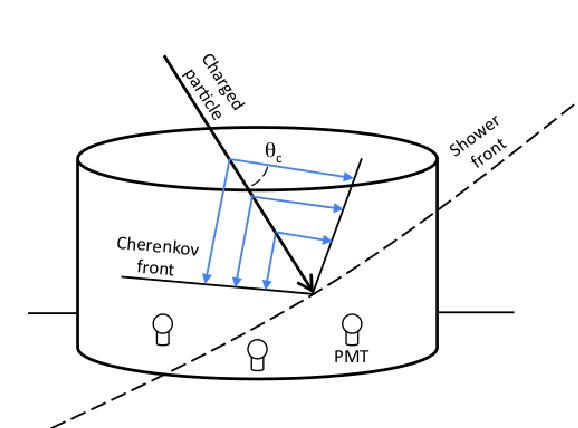}
    \caption{ Water Cherenkov Technique : The secondary charged particles present in the EAS will produce Cherenkov photons inside water tanks, whcih are detected by PMTs located at the bottom of each tank. Image courtesy : HAWC collaboration.}
    \label{fig:wc}
\end{figure}

In Water Cherenkov Technique (WCT), an array of PMTs is placed inside a water tank or a pool. These PMTs detect Cherenkov radiation emitted in the water by the secondary charged particles present in EAS (see figure \ref{fig:wc}). In water, the  Cherenkov emission angle is large ($\sim 41^\circ$), hence PMTs can be placed with larger spacings between them.  
The Multiple Institution Los Alamos Gamma-Ray Observatory (MILAGRO) detector \cite{milagro1,milagro2} built following WCT, consisting of a large water pool of area
60 m $\times$ 80 m and depth of 8 m, was located in New Mexico at an altitude of 2630 m. To prevent contamination from the ambient light entire pool was covered with a light-tight cover. There were 723 PMTs of 20 cm diameter arranged in two layers, with 450 PMTs in the top layer at a depth of 2.5m and remaining PMTs at the bottom of the water reservoir. The top layer was used to tag muons. Muon tagging is necessary to discriminate $\gamma$-ray showers from cosmic ray showers. the muon content of $\gamma$-ray initiated showers is negligible as a cross-section for photo-pion production is very small compared to cross-section for pair-production. On the other hand, 
hadron-initiated showers have a large number of muons, which survive till the  observational level.

\begin{figure}[h]
    \centering
    \includegraphics[width=0.65\textwidth]{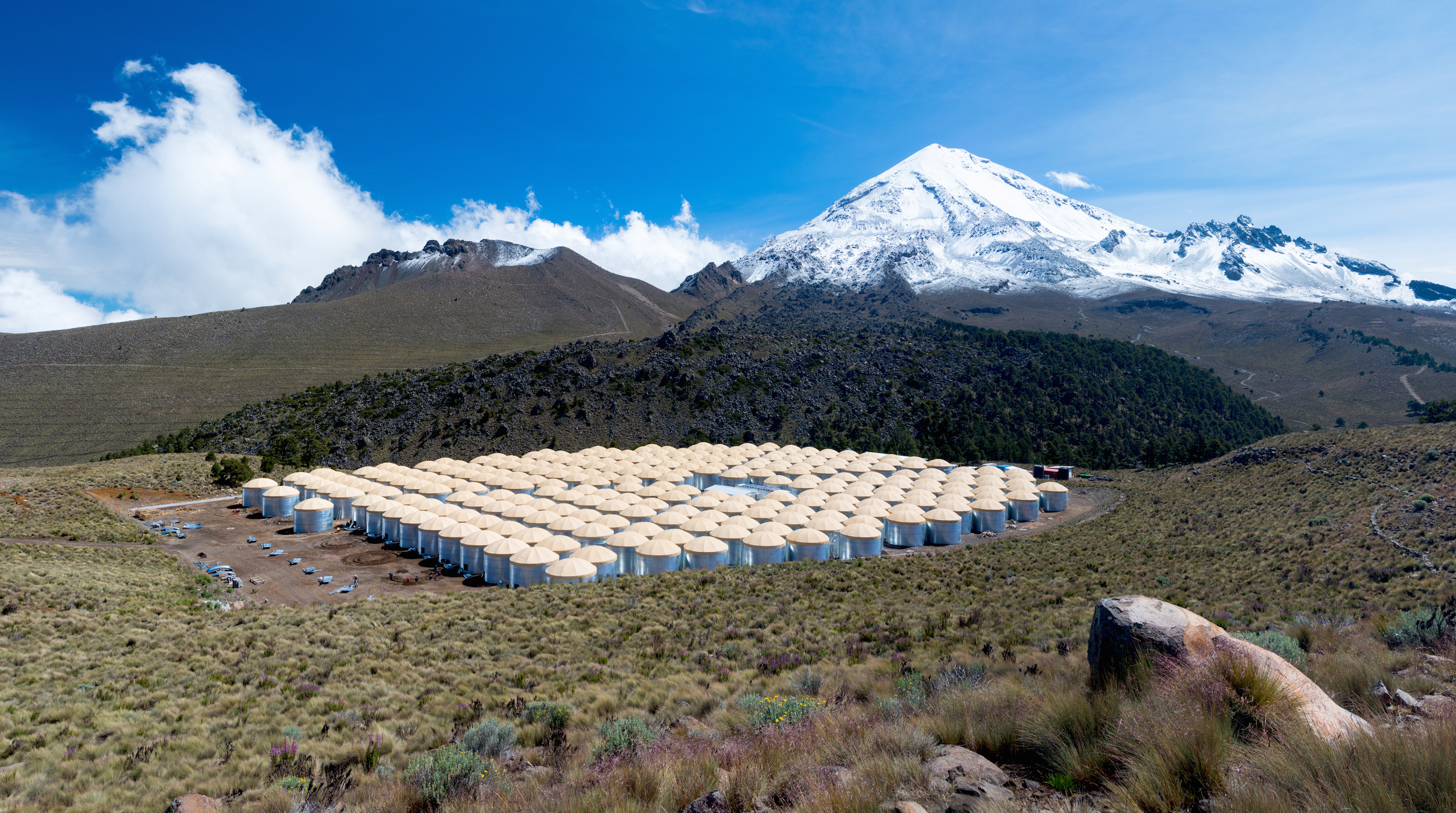}
    \caption{The High Altitude Water Cherenkov (HAWC) Experiment. Picture courtesy : HAWC collaboration.}
    \label{fig:hawc}
\end{figure}

Based on the experience gained with MILAGRO, the High-Altitude Water Cherenkov\footnote{https://www.hawc-observatory.org/} (HAWC) experiment is built, which is located on the Sierra Negra volcano in Mexico at an altitude of 4100 m (see figure \ref{fig:hawc}) \cite{hawc1,hawc2,hawc3}. There are 300 steel tanks, 5.4 m in height and 7.3 m in diameter, filled with purified water, spread over an area of 22000 ${m^2}$. Each tank contains a light-tight plastic bladder, which is filled with water to a depth of 4.5 m, ensuring particles in the shower are fully absorbed. There are four upward-facing 8-inch PMTs located at the bottom of each tank. Electrons and positrons present in the shower produce Cherenkov photons in the water which are detected by the PMTs. Muons present in the EAS initiated by cosmic rays  emit Cherenkov radiation along their path inside the tank before exiting. These muons  usually pass closer to one PMT than to the rest, thereby producing an excess signal in one PMT, which can be used to discriminate cosmic ray induced events from $\gamma$-ray induced events. HAWC has carried out the most sensitive sky survey with FoV $>$ 1.5 sr and with a duty cycle of more than 90\% at TeV energies till date. It has detected $\gamma$-rays above 100 TeV from some galactic sources \cite{hawc-pev}.

\subsection{Particle Detector Arrays}
The Extensive Air Shower Arrays use solid-state detectors to detect charged secondaries present in EAS. One such array called Tibet air shower array or Tibet AS$\gamma$
is located at Yangbajing in Tibet, China at an altitude of 4300 m a.s.l.
It is operational since 1990 and went through gradual upgradations over
several years. From the configuration of 533 fast timing (FT) plastic scintillators
of 0.5 $m^2$ area each, spread over 22050 $m^2$ in 1999, it was upgraded to
761 FT scintillator counters and another 28 density (D) counters with
effective area of 36900 $m^2$ in December 2003 \cite{Amenomori_2015}.
Later water Cherenkov muon detectors
with an area of $\sim$ 3400 $m^2$ were added to this array by 2014 \cite{Amenomori_2019}.
This array has an angular resolution of about 0.2$^\circ$
at 100 TeV. $\gamma$-ray event selection is based on direction reconstruction
and consequent rejection of isotropic cosmic ray background for air
shower array with only scintillators. Using this method signal from
Crab nebula was successfully detected at energies above 3 TeV
\cite{Amenomori_1999}. This was the first clear detection of $\gamma-$ray signal from a point source using an air shower array. This was
followed by successful detection of blazars Mrk 501 
and Mrk 421. 
After incorporating water Cherenkov muon
detectors in the array, based on measured muon numbers in the shower, cosmic
ray rejection was improved to 99.92\% with energies $>$ 100 TeV
leading to the first detection of photons with $E$ $>$ 100 TeV
from Crab nebula \cite{Amenomori_2019}.

There is another air shower array located at Yangbajing, 
Astrophysical Radiation with Ground-based Observatory at YangBaJing (ARGO-YBJ).
It consists of a single layer of Resistive Plate Counters (RPCs),
operated in streamer mode and with a modular structure. The basic
module is a cluster covering an area of 5.7 m $\times$ 7.6 m composed
of 12 RPCs, each with an area of 1.23 m $\times$ 2.85 m. There are
130 clusters in the central carpet of 74 m $\times$ 78 m with an
active area of $\sim$ 93\%. This central carpet is surrounded by
23 additional carpets or guard rings. The total area of the array is
110 m $\times$ 100 m \cite{Aielli_2010,Aielli_2006}. 
Using arrival time information of shower at various RPCs,
the position of shower core and arrival direction of primary is
reconstructed. Several sources like Crab, Mrk 421 have been detected
at energies above 300 GeV successfully \cite{Aielli_2010}.

Based on these previous experiences with ARGO-YBJ,  the Large High Altitude Air Shower Observatory (LHAASO) is being established in China \cite{lhaaso1,lhaaso2,lhaaso3}. 
LHASSO is under installation at a high altitude (4410m) location at the Daochen site,
Sichuan Province.  This experiment is designed with multiple
detecting methods, to carry out 3D observations of the EAS and is expected to have much better sensitivity 
in comparison with the previous similar experiments.

\begin{figure}[h]
    \centering
    \includegraphics[width=0.65\textwidth]{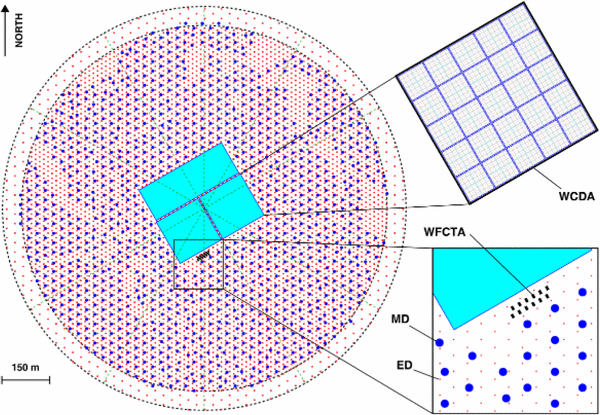}
    \caption{The layout of the LHAASO observatory. Multiple components of this experiment are shown separately. Image courtesy : LHASSO collaboration.}
    \label{fig:lhasso}
\end{figure}

\begin{figure}[h]
    \centering
    \includegraphics[width=0.75\textwidth]{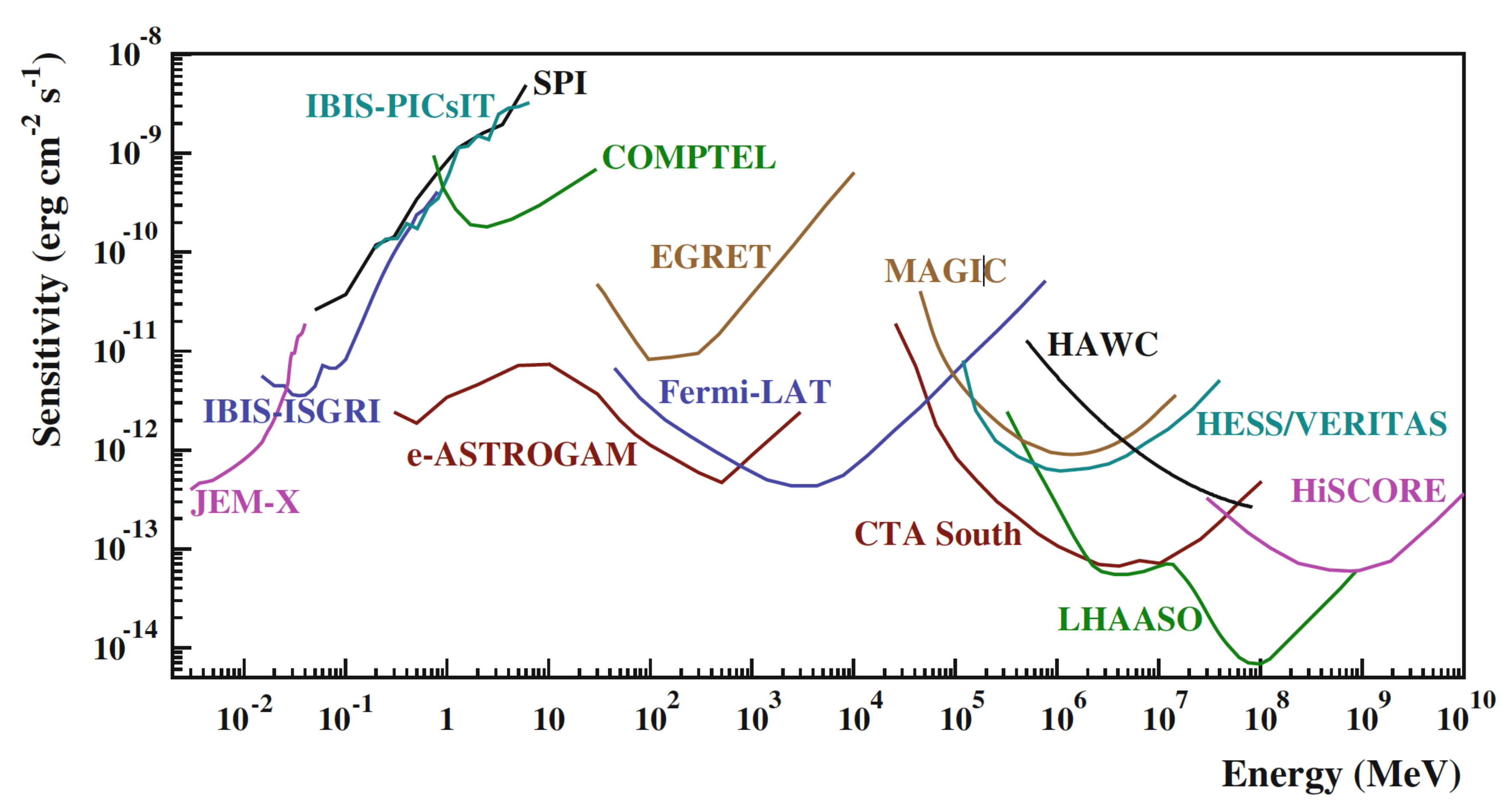}
    \caption{Point source differential sensitivity curves for different x-ray and $\gamma$-ray instruments. For IACTs e.g., MAGIC, VERITAS, H.E.S.S. \& CTA sensitivities are obtained for 50 hours. For HAWC, sensitivity is estimated for 5 years and for LHASSO 1 year. Figure is reproduced with permission from \cite{sensitivity-all}.}
    \label{fig:sensitivity}
\end{figure}

The LHAASO experiment mainly includes three detector sub-arrays; 1.3 km$^2$ array (KM2A), the central part is the Water Cherenkov light Detector Array (WCDA) consists of three ponds with a total area of 78000 $m^2$; there is also a telescope array (WFCTA), which is composed of 18 Wide Field Cherenkov light Telescopes (WFCTs) (see figure \ref{fig:lhasso}). The whole KM2A array  consists of 5195 Electromagnetic Detectors (ED), 1 $m^2$ each and 1188 Muon
Detectors (MD), 36 $m^2$ each. The KM2A can identify $\gamma$-ray showers from cosmic ray showers using MDs. At energies around 1 PeV it has unprecedented rejection power of $\sim 10^{-5}$. KM2A detects $\gamma$-rays with energies $>0.1PeV$. The energy resolution and angular resolution of KM2A are $<$20\% and $0.25^{\circ}$ respectively. WCDA can detect low energy (0.1TeV) $\gamma$-rays compared to KM2A. WCDA is ideal to carry out deep surveys for high energy sources. WFCTA detects Cherenkov light emitted by air showers produced by cosmic rays and $\gamma$-rays. These telescopes have wide field of view $32^{\circ}$ $\times$ $112^{\circ}$.
The current sensitivity of LHASSO is already an order of magnitude better compared to ACTs in the energy range of 100 TeV and above \cite{lhasso3}, 
see figure~\ref{fig:sensitivity}. In its early years of operation, LHASSO has detected several candidate PeVatrons in the galactic plane with high 
significance\cite{lhasso-nature}.

\begin{figure}[h]
    \centering
    \includegraphics[width=0.65\textwidth]{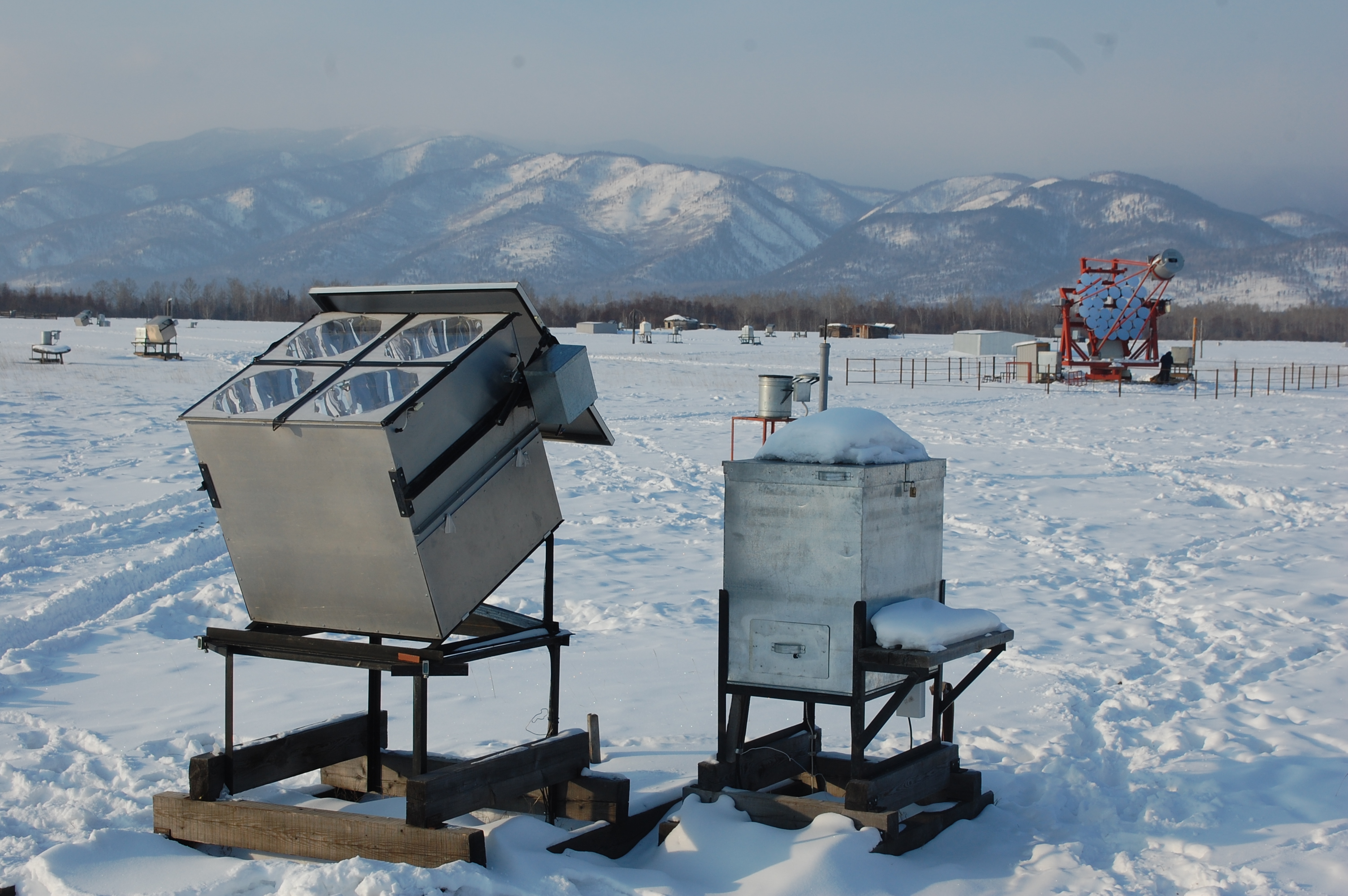}
    \caption{One of the TAIGA-HiSCORE integrating Cherenkov detector stations with a wide Field of View is seen at the front in the above photograph. Each detector station consists of 4 large area PMTs. One installed TAIGA-IACT telescope seen at the back. Picture courtesy : TAIGA Collaboration.}
    \label{fig:TAIGA}
\end{figure}

The new hybrid multi-component Tunka Advanced Instrument for cosmic ray physics and Gamma Astronomy (TAIGA) experiment aims to detect air showers induced by $\gamma$-rays and cosmic rays.
The TAIGA experiment is located at the Tunka valley (altitude 650 m), 50 km West of Lake Baikal, Russia.
TAIGA-IACT and TAIGA-HiSCORE  will be used to detect VHE $\gamma$-rays above 30 TeV (see figure \ref{fig:TAIGA}). TAIGA-IACT \cite{taigaiact} will be an array of 16 IACTs. The reflector area of each telescope will be $10m^2$ and the imaging camera will have 560 PMTs. TAIGA-HiSCORE (Hundred Square km Cosmic ORigin Explorer) \cite{taigahiscore} will be  an array of integrating Cherenkov detectors with a wide field of view (0.6 sr). Each detector station will consist of 4 large area PMTs. There will be 100 such detectors (with 150-200m detector spacing) covering an area of $\sim 5~km^2$. 


All the air shower arrays mentioned above, LHASSO, HAWC and TAIGA, even with their  wide field of view, cannot access sources located in the southern hemisphere, which include the galactic centre. Hence, to achieve full sky coverage, a southern array is under consideration. Southern Wide-field Gamma-ray Observatory (SWGO)\footnote{https://www.swgo.org/SWGOWiki/doku.php} is proposed to be built in the Andes at an altitude of 4.4 km a.s.l. \cite{swgo}.
The proposed array will consist of water Cherenkov detectors aided by muon detectors for better gamma-hadron separation. 


\section{Conclusion}
\label{con}
Very High Energy $\gamma$-ray astronomy has undergone a remarkable transformation in the last 3-4 decades. Initially, progress was very slow, first attempts were made in the 1960s but astronomers had to wait for almost three decades before the first source was detected in the 1980s by Whipple $\gamma$-ray telescope. 
Whipple telescope proved that the imaging technique is the most powerful method to identify $\gamma$-ray initiated showers from overwhelming background of cosmic rays. Following the footsteps of Whipple, the current generation of telescopes, H.E.S.S., MAGIC and VERITAS have revolutionised VHE $\gamma-$ray astronomy. In the beginning of this century, only a handful of sources were detected by ground-based telescopes and at present, the number of sources detected is more than two hundred. It is expected that the number of sources detected by the upcoming CTA will be few thousand. On the other hand, Air Shower Array experiments like HAWC and LHASSO will carry out extensive surveys at Ultra High Energy regime and with higher sensitivities above 100 TeV. Ground-based $\gamma$-ray astronomy is now entering a fascinating phase and is going to play a very important role in multi-messenger astronomy in the coming decades.

\section{Acknowledgement}
DB acknowledges Science and Engineering Research Board - Department of Science and Technology for Ramanujan Fellowship - SB/S2/ RJN-038/2017. PM acknowledges the generous support of the Stanislaw Ulam fellowship (PPN/ULM/2019/1/00096/A/00001) by Polish National Agency for Academic Exchange (NAWA). VRC acknowledges the support of the Department of Atomic
Energy, Government of India, under Project Identification No. RTI4002. 

%
%

\end{document}